\documentclass[traditabstract,longauth]{aa}  
\usepackage{graphicx}
\usepackage{txfonts}
\usepackage{natbib}
\bibpunct{(}{)}{;}{a}{}{,}
\begin{document}
   \title{\textit{Herschel}\thanks{{\it \textit{Herschel}} is an ESA space observatory with science instruments provided by European-led Principal Investigator consortia and with important participation from NASA.} observations of B1-bS and B1-bN: two first hydrostatic core candidates in the Perseus star-forming cloud}
\titlerunning{B1-bS and B1-bN: two first hydrostatic cores candidates in the Perseus star-forming cloud}

   \author{Stefano Pezzuto\inst{1}\and D. Elia\inst{1}\and E. Schisano\inst{1}\and F. Strafella\inst{2}\and J. Di Francesco\inst{3,4}\and S. Sadavoy\inst{3,4}\and P. Andr\'e\inst{5}\and M. Benedettini\inst{1}\and J.P. Bernard\inst{6}\and A.M. di Giorgio\inst{1}\and A. Facchini\inst{1}\and M. Hennemann\inst{5}\and T. Hill\inst{5}\and V. K\"onyves\inst{5}%\and A. Men`shchikov\inst{5}
\and S. Molinari\inst{1}\and F. Motte\inst{5}\and Q. Nguyen-Luong\inst{7,5}\and N. Peretto\inst{5}\and M. Pestalozzi\inst{1}\and D. Polychroni\inst{1}\and K.L.J. Rygl\inst{1}\and P. Saraceno\inst{1}\and N. Schneider\inst{8,5}\and L. Spinoglio\inst{1}\and L. Testi\inst{9}\and D. Ward-Thompson\inst{10}\and G.J. White\inst{11,12}
          }

   \institute{IAPS - INAF, via Fosso del Cavaliere, 100, I-00133 Roma, Italy\\
              \email{pezzuto@iaps.inaf.it}
   \and Dipartimento di Fisica, Universit\`a del Salento, CP 193, I-73100 Lecce, Italy
   \and National Research Council Canada, Herzberg Institute of Astrophysics, 5071 West Saanich Road, Victoria BC Canada, V9E 2E7
   \and Department of Physics and Astronomy, University of Victoria, PO Box 355, STN CSC, Victoria BC Canada, V8W 3P6
   \and Laboratoire AIM, CEA/DSM-CNRS-Universit\'e Paris Diderot, IRFU/Service d'Astrophysique, CEA Saclay, 91191 Gif-sur-Yvette, France
   \and CESR, Observatoire Midi-Pyr\'en\'ees (CNRS-UPS), Universit\'e de Toulouse, BP 44346, 31028 Toulouse, France
   \and Canadian Institute for Theoretical Astrophysics, University of Toronto, 60 St. George Street, Toronto, ON M5S 3H8, Canada
   \and LAB/OASU-UMR5804, CNRS- University of Bordeaux, 33270 Floirac, France
   \and ESO, Karl Schwarzschild-Strasse 2, 85748 Garching bei M\"unchen, Germany
   \and Jeremiah Horrocks Institute, University of Central Lancashire, PR1 2HE, UK
   \and Department of Physics and Astrophysics, Open University, Walton Hall, Milton Keynes MK7 6AA, UK
   \and 14 Rutherford Appleton Laboratory, Chilton OX11 0QX, UK}
   \date{}

% \abstract{}{}{}{}{} 
% 5 {} token are mandatory
 
  \abstract
  % context heading (optional)
   {We report far-IR \textit{Herschel} observations obtained between 70~$\mu$m and 500~$\mu$m of two star-forming dusty condensations, B1-bS and B1-bN, in the B1 region of the Perseus star-forming cloud. In the Western part of the Perseus cloud, B1-bS is the only source detected in all of the 6 PACS and SPIRE photometric bands without being visible in the \textit{Spitzer} map at 24~$\mu$m. B1-bN is clearly detected between 100~$\mu$m and $250$~$\mu$m. We have fitted the spectral energy distributions of these sources to derive their physical properties, and find that a simple greybody model fails to reproduce the observed SEDs. At least a two-component model, consisting of a central source surrounded by a dusty envelope is required. The properties derived from the fit, however, suggest that the central source is not a Class~0 object. We then conclude that while B1-bS and B1-bN appear to be more evolved than a pre-stellar core, the best-fit models suggest that their central objects are younger than a Class~0 
source. Hence, they may be good candidates to be examples of the first hydrostatic core phase. The projected distance between B1-bS and B1-bN is a few Jeans lengths. If their physical separation is close to this value, this pair would allow the mutual interactions between two forming stars at a very early stage of their evolution to be studied.}% In the past, the very young nature of B1-bS has not been fully recognized because of its closeness to the more evolved \textit{Spitzer} source S295. \textit{Herschel} data clearly show that these two sources are spatially separate.}% Therefore, it is possibile that B1-bS/bN and Bolo~81 are also two different objects.}% If high spatial resolution observations in the millimetre confirm the young nature of these sources, they will offer the possibility not only to observe the very first stages of star formation which, lasting only few $10^2$--$10^3$ years, are the most difficult to observe, but also the mutual interactions between two forming stars, given the very small separation of 20\arcsec, or 4700~AU at 235~pc.}

%   \keywords{Stars: protostars - Star: individual: [HKM99] B1-bS - Star: individual: [HKM99] B1-bN - Star: individual: [HRF2005] 2 - Star: individual: [HRF2005] 3 - Star: individual: [EYG2006] Bolo 81 - Star: individual: [EDJ2009] 295 }

   \maketitle
%
%________________________________________________________________

\section{Introduction}
During the formation of a low-mass star, the first hydrostatic core (FHSC) phase is one of the shortest phases, lasting only $10^2$--$10^3$~yrs, depending on how the collapse proceeds \citep{bate}. The FHSC phase is characterized by the presence of a central object with the mass of a giant planet and having a size of just a few astronomical units. This phase lasts until the central object temperature reaches $\sim$2000~K. At that point, the molecular hydrogen dissociates, causing a second collapse that brings the central object to the typical size of a protostar, starting the Class~0 phase \citep{class0}. The short duration of the FHSC phase makes it difficult to observe an FHSC in a star-forming region.% Indeed, up to now only few objects, see below, have been proposed as \textbf{FHSC} candidates.

Before the FHSC phase, the star-forming core (a ``pre-stellar core'') has a spectral energy distribution (SED) that can be modelled as a greybody spectrum. After this FHSC phase, the SED in the far-infrared (FIR) still resembles a greybody spectrum, but the emission of the central forming protostar starts to be visible at shorter wavelengths, i.e., $\lambda < 70$~$\mu$m. Theoretical models \citep[e.g.,][]{omukai,sato} show that the SED deviates from a pure greybody shape during the FHSC phase at wavelengths $\lambda\la200$~$\mu$m. At even shorter wavelengths, e.g., for $\lambda\la30$~$\mu$m, the emission is still too faint to be detected. Taking into account that the photometric bands ot the \textit{Spitzer} MIPS instrument \citep{MIPS} were centred in the FIR at 24~$\mu$m, 70~$\mu$m, and 160~$\mu$m, the expected signature for a 1~$M_\odot$ FHSC SED includes a detection at 70~$\mu$m and the non-detection at 24~$\mu$m \citep{comm}.

None of the FHSC candidates meets these photometric requirements. In the Perseus star forming region, the candidate Per~58 \citep{fhsc,dunham} is visible in the 24~$\mu$m \textit{Spitzer} map. Similarly, the candidate MMSI in Chamaeleon \citep{belloche,cordinor} is observed both at 24~$\mu$m and at 70~$\mu$m. The proposed FHSC candidates L1451-mm \citep{pineda}, Bolo~45 \citep{schnee}, and CB~17 \citep{chenF}, are indeed not visible at 24~$\mu$m, but they remain undetected also at 70~$\mu$m in the \textit{Spitzer} maps, making the SED-based classification less firm. Finally, L1448-IRS2E \citep{chenF2} was not even detected at 160~$\mu$m. It must be said, however, that the knowledge of the SED alone is not enough to firmly identify an FHSC. Spectral lines and interferometric observations are necessary and, indeed, these FHSC candidates were not proposed based on continuum data only.

Further, the (non)detection in a certain band clearly depends on the distance to the source. To predict the expected flux from an FHSC, the distance of the star forming region in Taurus (150~pc) is often taken as reference distance \citep[e.g.,][]{tomida,comm}. Moving to larger distance, however, makes stronger the requirement on non detection at 24~$\mu$m, while the detection at 70~$\mu$m may become less stringent. In fact, the non-detection at 70~$\mu$m has been explained by \citet{fhsc} on the basis that an FHSC could, in principle, be detected with \textit{Spitzer}, but the ``Cores to Disks'' (c2d) survey \citep{evans} was not sufficiently sensitive to detect these sources. The recent launch of the \textit{Herschel} Space Observatory satellite \citep{HSO} has now opened up the possibility to observe in the FIR bands between 70~$\mu$m and 500~$\mu$m, with unprecedented sensitivity and spatial resolution. Among the observing programs, the Herschel Gould Belt survey \citep[GBS;][]{pandreGB} aims to 
obtain a complete census of pre-stellar cores and Class~0 sources in the closest star-forming regions. Since FHSCs can be considered extremely young Class~0 sources, we expect to find new FHSC candidates among the numerous objects discovered with \textit{Herschel}.

One of the targets clouds of the GBS is the star-forming region in Perseus molecular cloud, at an average distance of $\sim$235~pc \citep{hirota}. Perseus hosts low and intermediate-mass young stellar objects, and so it is roughly between a low-mass star-forming cloud like Taurus, and a high-mass star-forming cloud like Orion.

As a starting point for our analysis of these observations, we looked for sources detected in the \textit{Herschel} map at 70~$\mu$m without any counterpart in the corresponding \textit{Spitzer} 24~$\mu$m map.
%This cloud has been observed in a large number of photometric bands and spectral lines: being part of the \textit{Spitzer} ``Cores to Disks'' (c2d) program \citep{evans} it has been observed with the InfraRed Array Camera (IRAC) \citep{jorge} and the Multiband Imaging Photometer (MIPS) \citep{rebull}; it has also been mapped in the sub-mm \citep{hatchell,kirk} and at 1.1~mm \citep{enoch}.
We found one such source and in this paper we discuss the physical properties that can be derived from the analysis of its SED. This object is spatially associated with the source B1-bS discovered by \citet{hirano}. They found this source through interferometric observations carried out with the Nobeyana Millimeter Array (NMA) at 3~mm. They also found a second source, B1-bN, $\sim$20\arcsec\ north of B1-bS, clearly visible in the \textit{Herschel} PACS maps at $\lambda\ge100$~$\mu$m. Combining the NMA data with other single dish observations between 350~$\mu$m and 2~mm, and spectral observations of H$^{13}$CO$^+$, \citet{hirano} concluded that these two sources were younger than already known Class~0 objects. Their observations, however, were all taken in the Rayleigh-Jeans part of their SEDs, causing a large uncertainty in the determination of the temperature and, in turn, of the mass. \textit{Herschel} data now extend the knowledge of the SEDs exactly where the peak of the emission falls.

On the basis of \textit{Herschel} fluxes, which are incompatible with a greybody, we reconsider the physical properties of these two sources, and we propose they are new promising FHSC candidates. The rest of the paper is so organized: in Section~2, the observations are presented and the algorithm of source detection, CuTEx, is briefly described. In Section~3, we fit the SEDs with a two-component model that describes the FHSC emission in a simple way. We show that alternative approaches fail to reproduce the SEDs. In Section~4, we discuss the results, and, in Section~5, our work is summarized. Appendix~A gives a detailed description of the SED fitting, including ancillary explanations on the colour corrections and on the Gaussian fitting of sources detected with \textit{Herschel}. We also present a new formula to derive the mass of an object whose SED is fitted with a greybody. Appendix~B describes an alternative approach to the sigma-clipping algorithm for outlier detection, where the threshold for the 
detection depends on the size of the sample.

%The paper is organized as follows: in Section~2 we give an overview of the observations of Perseus and present our data reduction and analysis; in Section~3 we present the results of the SED fitting. In Section~4 we summarize our work. Appendix~A describes a sigma-clipping method to detect outliers with a variable threshold; in Appendix~B we detail our fitting procedure and we derive a new formula to compute the mass of a greybody.

\section{Observations and data analysis}
The observations of Perseus cover a total area of about 10~deg$^2$, mapped with the PACS \citep[Photodetector Array Camera \& Spectrometer;][]{PACS} and SPIRE \citep[Spectral and Photometric Imaging Receiver;][]{SPIRE,bruce} instruments using the parallel mode with the telescope scanned at 60\arcsec/s. In this observing mode, the two instruments observe almost the same field with SPIRE delivering data in its three bands (250~$\mu$m, 350~$\mu$m, and 500~$\mu$m), and PACS in two out of three bands, 70~$\mu$m and 160~$\mu$m in our case. A smaller area, 6.7~deg$^2$, was mapped again with PACS at 100~$\mu$m and 160~$\mu$m only, with the telescope scanned at 20\arcsec/s, to reach a higher spatial resolution and a better sensitivity.

The observations we consider in this paper cover a region of about 5.6~deg$^2$, centred on NGC~1333. Table~\ref{log} summarizes the observations. Preliminary results from the parallel mode observations were recently presented by \citet{sarah2}, % who studied a small, high-extinction region ($A_\mathrm{V}>5$~mag) east of B1. Another work 
\ and by Bressert et al. (submitted).% analysed the spatial correlation of the sources in the Perseus and Serpens clouds.
\ A complete description of this region based on \textit{Herschel} data will be given in an upcoming paper (Pezzuto et al., in preparation).

\begin{table*}
\caption[]{The log of the obervations. Column OBSID reports the Observational Identifier which specifies an observation in the Herschel Science Archive. Each field was observed twice along two almost orthogonal directions to better remove the instrumental $1/f$ noise. Date refers to the start of the observation; OD is the operational day (OD 1 is 14 May 2009). Centre gives the central coordinates of the map. Size is in arcminutes. The Observing mode column reports Parallel, for PACS and SPIRE data obtained in parallel mode, or PACS for data obtained with PACS only. PACS bands column reports the effective wavelength of the selected PACS filters. Speed is the scanning velocity of the telescope.\label{log}}
\begin{tabular}{cccccccc}
\hline
OBSID&Date&OD&Centre&Size&Obs. mode&PACS bands&Speed\\\hline
1342190326&09/02/2010&271&3$^\mathrm{h}$29$^\mathrm{m}$39$^\mathrm{s}$ +30$^\mathrm{d}$54\arcmin32\arcsec&135\arcmin$\times$150\arcmin&Parallel&70~$\mu$m, 160~$\mu$m&60\arcsec/s\\
1342190327&09/02/2010&271&3$^\mathrm{h}$29$^\mathrm{m}$41$^\mathrm{s}$ +30$^\mathrm{d}$53\arcmin58\arcsec&150\arcmin$\times$135\arcmin&Parallel&70~$\mu$m, 160~$\mu$m&60\arcsec/s\\\hline
1342227103&22/08/2011&831&3$^\mathrm{h}$30$^\mathrm{m}$17$^\mathrm{s}$ +30$^\mathrm{d}$48\arcmin05\arcsec&135\arcmin$\times$135\arcmin&PACS&100~$\mu$m, 160~$\mu$m&20\arcsec/s\\
1342227104&23/08/2011&831&3$^\mathrm{h}$30$^\mathrm{m}$17$^\mathrm{s}$ +30$^\mathrm{d}$47\arcmin59\arcsec&135\arcmin$\times$135\arcmin&PACS&100~$\mu$m, 160~$\mu$m&20\arcsec/s\\\hline
\end{tabular}
\end{table*}

Data were processed with version 8.0.1559 of HIPE \citep{ott}, except: a) deglitching, which was performed with a sigma-clipping algorithm with a variable threshold detailed in Appendix~B, and b) map reconstruction, which was made using the ROMAGAL code \citep{alessio}. The zero point of the flux calibration for extended emission was established by comparing \textit{Herschel} data with \textit{Planck} and IRAS data over the same area \citep{jpb}. The result of the comparison is a set of offsets, one for each band, algebrically added to the observed images. For photometric measurements, however, where a background is estimated and subtracted from the source flux, as in this work, the zero points are actually not important.

%The final maps are shown in Fig.~\ref{maps}: the 6 bands have been combined in two RGB figures, one with PACS and one with SPIRE data.% The noise in the PACS maps is (in MJy/sr): 14.1 (70~$\mu$m), 7.0 (100~$\mu$m), 6.5 (160~$\mu$m scan speed 20"), 7.9 (160~$\mu$m scan speed 60"); for SPIRE maps: 2.9, 1.8, 1.0, for 250, 350 and 500~$\mu$m, respectively. These latter values may overestimate the instrumental rms since it is nearly impossible to find regions free of diffuse emission in SPIRE maps.

%\onlfig{1}{
%\begin{figure*}
%\includegraphics[scale=0.3]{PACSrgb.jpg}
%\includegraphics[scale=0.3]{SPIRErgb.jpg}
%\caption{fare PACS/SPIRE? The region of NGC~1333 in Perseus observed with \textit{Herschel}. Left: rgb map with PACS 70~$\mu$m (blue), 100~$\mu$m (green) and 160~$\mu$m (red); right: rgb map with SPIRE 250~$\mu$m (blue), 350~$\mu$m (green) and 500~$\mu$m (red). The green (PACS) or magenta (SPIRE) small box shows the position of the 1.5~arcmin$^2$ region containing the sources B1-bS and B1-bN (see the figure below). The white box displays the part of B1 discussed by \citet{sarah2}. The 100~$\mu$m map has been observed with PACS only with a scan speed of 20\arcsec/s, all the other bands with PACS/SPIRE in parallel mode at a scan speed of 60\arcsec/s. NGC~1333 is the bright region at $3^\mathrm{h}28^\mathrm{m}$ $31\fdg5$. The scale is in MJy/sr, stretched from 0 to 500 for all bands.\label{maps}}
%\end{figure*}}

\subsection{Source detection and photometry}
%Maps in the far infrared are very demanding in terms of data analysis. Source detection and photometry is a complex matter due to the presence of the diffuse emission which contributes with a strong signal spatially variable at all scales. % For this reason the Gould Belt consortium is working to define an optimal strategy for compact sources analysis. Many algorithms have been proposed to tackle this problem, see \citet{sasha} for details.
To identify sources in the field, we adopted the CuTEx algorithm \citep{derivatives}. This method computes the second derivative of the input image along four different directions. In the ``curvature'' image so obtained, we look for pixels where the second derivative is negative and its absolute value is above an user defined threshold with respect to the local fluctuactions. Isolated high-curvature pixels are rejected, and only groups of contiguous pixels, whose number depends on the sampling of the beam, are further considered. The curvature in each group is analyzed to verify the statistically significance of the local peak. The detection is done independently at each wavelength. Once a preliminary list of candidate sources is derived, photometry is recovered by fitting a 2D Gaussian profile. The position of the peaks of the candidate sources and their sizes, as derived from the curvature map, are used as initial guesses for the parameters of the 2D Gaussian. If two or more sources are closer than twice 
the width of the point spread function (PSF) at a given wavelength, their Gaussians are fitted simultaneously.

For $\lambda\ge160$~$\mu$m, the diffuse emission contributes a strong, variable signal, across all spatial scales. An important issue is then how to estimate the contribution of the diffuse background at the source position. CuTEx models such emission by fitting a linear 2D-polynomial (i.e., a plane) inside a subregion centred on each source, of size 6 times the PSF at the specific wavelength. The plane and the Gaussian(s) are fitted simultaneously. There is, however, no general consensus on how the background can be estimated: different approaches can give different results and one of the main, and to large extent unknown, uncertainties in the derived fluxes, and in turn on the shape of the SED, comes from the background subtraction. For this reason is important to compare the results of flux extraction with more than one method. To this aim, we also used \textit{getsources} \citep{sasha} to derive the fluxes of our sources.

Among the sources detected at 70~$\mu$m, we found that only one was not detected in the corresponding \textit{Spitzer} 24~$\mu$m map. This source is one of a pair of very young objects first discovered at millimetre wavelengths by \citet{hirano} and dubbed B1-bS and B1-bN. Our 70~$\mu$m source corresponds positionally to B1-bS, while B1-bN is clearly detected between 100~$\mu$m and 250~$\mu$m. The separation between the two sources is about 20\arcsec, larger than the beam of PACS in all bands. It is also larger than the beam of SPIRE at 250~$\mu$m ($\sim$18\arcsec), but smaller than the beams at 350~$\mu$m and 500~$\mu$m ($\sim$25\arcsec and $\sim$36\arcsec, respectively)%\footnote{See Sibthorpe et al. ``SPIRE Beam Model Release Note'', version 1.1, available on-line, for the precise size of the beams.}
. We attempted to fit two Gaussians to the SPIRE data using the positions of B1-bS and B1-bN at 160~$\mu$m as initial guesses. During the fitting procedure the initial distance between the two objects was kept constant.

\textit{Herschel} data were complemented with archival \textit{Spitzer} data from the c2d survey \citep{evans}, SCUBA Legacy Catalogue data \citep{james} at 450~$\mu$m and 850~$\mu$m, and the Bolocam data \citep{enoch} at 1.1~mm. We ran CuTEx on all these maps except the \textit{Spitzer} ones (see below). Again, the positions of the sources at 160~$\mu$m were used as initial guesses. The effective FWHM resolutions of the SCUBA Legacy Catalogue data at 450~$\mu$m is 17\farcs3, so that the two sources can be considered to be fairly separated. At 850~$\mu$m, however, the effective FWHM is  22\farcs9 and encompasses both B1-bS and B1-bN, causing the photometry to be more uncertain. At 1.1~mm, the Bolocam beam of 31\arcsec makes it impossible to distinguish the two objects.

\begin{table}%\tiny
\caption[]{Photometric fluxes of B1-bS and B1-bN. The four rows denoted as IRAC$i$ report the 1\,$\sigma$ upper limits derived on the maps at the positions of the sources at 160~$\mu$m. Size is the instrumental FWHM. The row denoted as MIPS1 reports the 5\,$\sigma$ upper limit at 24~$\mu$m, found by analyzing the \textit{Spitzer} map with DAOPHOT inside an aperture of 12\arcsec. For all the other wavelengths, the fluxes were measured by fitting a 2D Gaussian at the positions found by CuTEx. Size is the FWHM of the elliptical Gaussian fitted to each source, circularized and deconvolved with the instrument beam size. The 70~$\mu$m flux of B1-bN is equal to that of a point source having a 1\,$\sigma$ rms peak flux, so we consider it as an upper limit. The 160~$\mu$m data are from the map scanned at 20\arcsec/s. The uncertainties were estimated as the differences in the fluxes extracted with two independent methods: CuTEx and \textit{getsources} \citep{sasha}. \textit{Herschel} fluxes have had Gaussian fit 
corrections applied but no colour corrections, see Appendix~A.1.\label{photData}}
\begin{tabular}{ll|lc|lc}%p{0.5\linewidth}l}
\hline
&&\multicolumn{2}{c}{\textbf{B1-bS}}&\multicolumn{2}{|c}{\textbf{B1-bN}}\\
%\noalign{\smallskip}
\multicolumn{1}{c}{$\lambda$}&\multicolumn{1}{c|}{Instr.}&\multicolumn{1}{c}{Flux}&Size&\multicolumn{1}{c}{Flux}&Size\\
\multicolumn{1}{c}{($\mu$m)}&&\multicolumn{1}{c}{(Jy)}&(\arcsec)&\multicolumn{1}{c}{(Jy)}&(\arcsec)\\
%\noalign{\smallskip}
\hline
%\noalign{\smallskip}
\phantom{00}3.8   &IRAC1&$<1.2\,10^{-6}$&1.66&$<1.0\,10^{-6}$&1.66\\
\phantom{00}4.5   &IRAC2&$<2.3\,10^{-6}$&1.72&$<1.6\,10^{-6}$&1.72\\
\phantom{00}5.8   &IRAC3&$<9.4\,10^{-6}$&1.88&$<6.0\,10^{-6}$&1.88\\
\phantom{00}8.0   &IRAC4&$<1.5\,10^{-5}$&1.98&$<7.7\,10^{-6}$&1.98\\
\phantom{00}24    &MIPS1&$<2.0\,10^{-4}$&12&$<2.0\,10^{-4}$&12\\
\phantom{00}70\tablefootmark{a}&PACS&\phantom{0}0.22$\pm$0.14&\phantom{0}4.4&\multicolumn{1}{c}{$<$0.050}&\phantom{0}6.9\\
\phantom{0}100\tablefootmark{a}&PACS&\phantom{0}2.29$\pm$0.26&\phantom{0}3.4 &\phantom{0}0.61$\pm$0.31&\phantom{0}8.5\\
\phantom{0}160\tablefootmark{a}&PACS&\phantom{0}9.1\phantom{0}$\pm$1.2&\phantom{0}5.8 &\phantom{0}3.24$\pm$0.80&\phantom{0}8.9\\
\phantom{0}250    &SPIRE&14.4\phantom{0}$\pm$1.4&10.4&\phantom{0}9.49$\pm$0.79&14.3\\
\phantom{0}350    &SPIRE&16.9\phantom{0}$\pm$4.8&17.8&12.7\phantom{0}$\pm$1.3&18.4\\
\phantom{0}450\tablefootmark{b}&SCUBA&19.1\phantom{0}$\pm$4.6&10.1&13.5\phantom{0}$\pm$3.7&10.2\\
\phantom{0}500    &SPIRE&15.8\phantom{0}$\pm$7.8&34.3&14.6\phantom{0}$\pm$4.9&34.3\\
\phantom{0}850\tablefootmark{b}&SCUBA&\phantom{0}3.8\phantom{0}$\pm$1.3&19.8&\phantom{0}3.3\phantom{0}$\pm$1.4&22.3\\
1100\tablefootmark{c}&Bolocam&\multicolumn{4}{c}{3.67$\pm$0.26~Jy within 59\farcs9$\times$40\farcs3}\\
%\noalign{\smallskip}
\hline
\end{tabular}
\tablefoot{\tiny \tablefoottext{a} Fluxes after the PSF correction discussed in Appendix~A; \tablefoottext{b} Size refers to the inner part of the SCUBA beam, see Appendix~A.2; \tablefoottext{c} \citet{enoch}: 2.61 Jy within 53\arcsec$\times$59\arcsec.}
\end{table}

\begin{figure*}[h]
\centering
\includegraphics[width=0.24\linewidth]{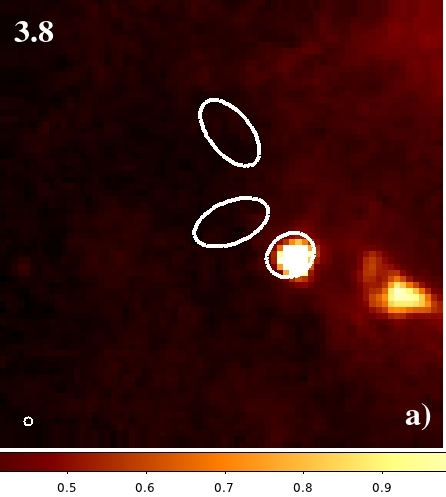}
\includegraphics[width=0.24\linewidth]{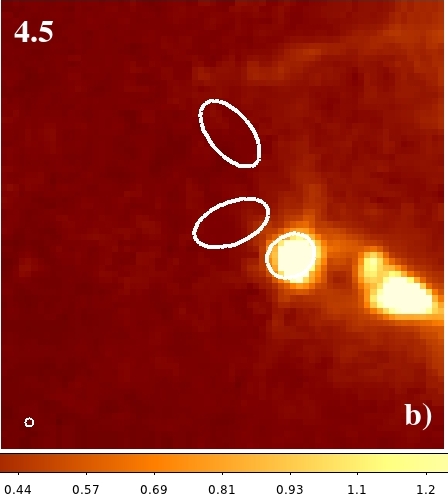}
\includegraphics[width=0.24\linewidth]{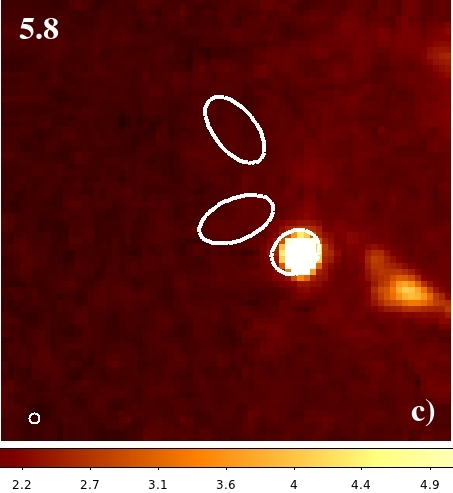}
\includegraphics[width=0.24\linewidth]{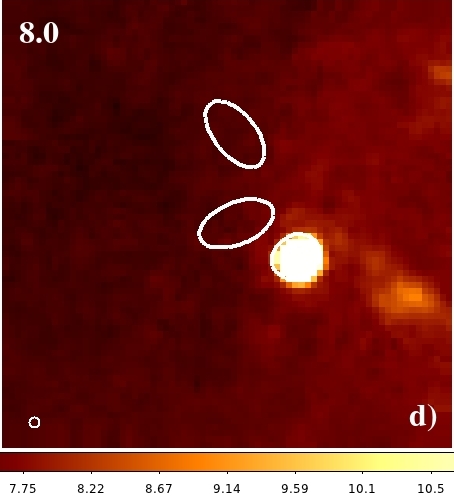}
\includegraphics[width=0.24\linewidth]{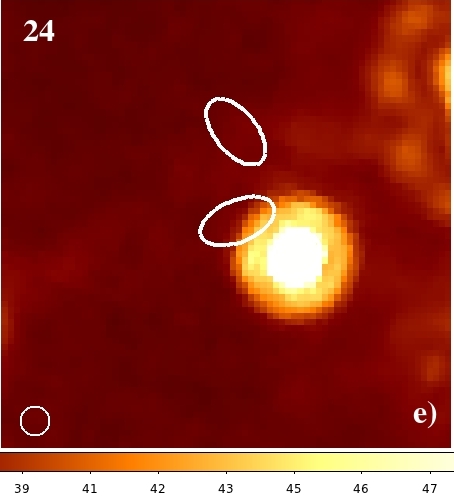}
\includegraphics[width=0.24\linewidth]{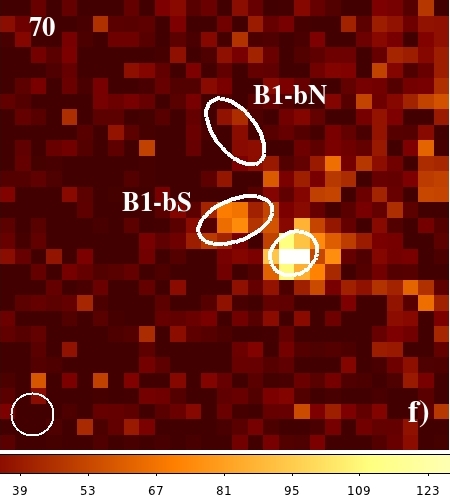}
\includegraphics[width=0.24\linewidth]{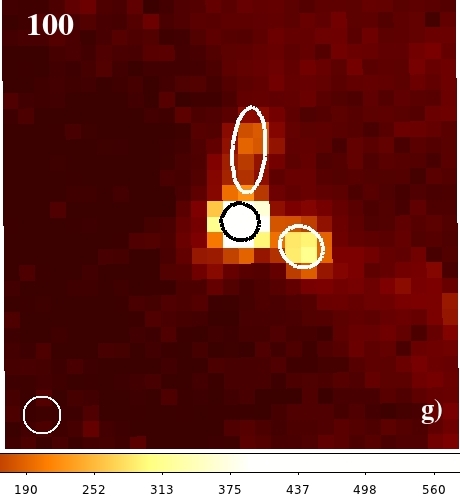}
\includegraphics[width=0.24\linewidth]{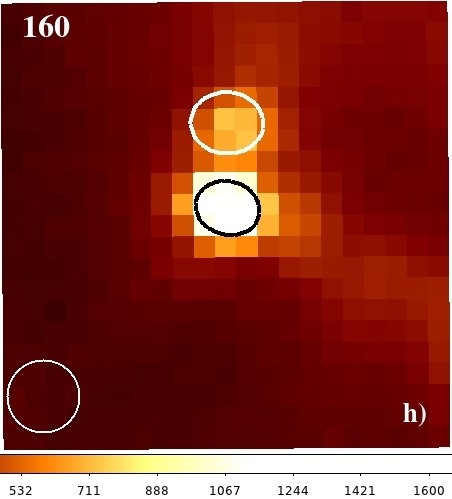}
\includegraphics[width=0.24\linewidth]{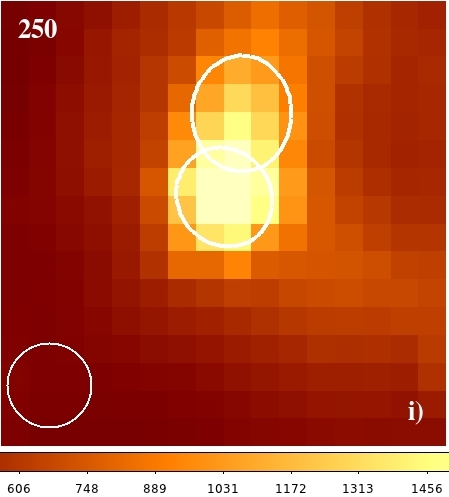}
\includegraphics[width=0.24\linewidth]{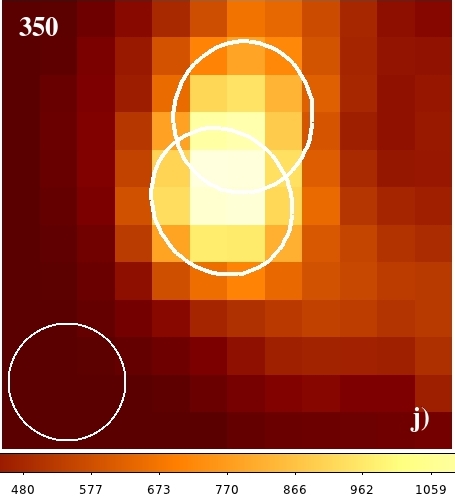}
\includegraphics[width=0.24\linewidth]{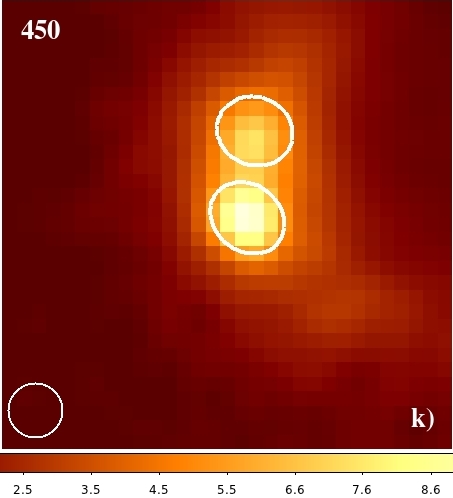}
\includegraphics[width=0.24\linewidth]{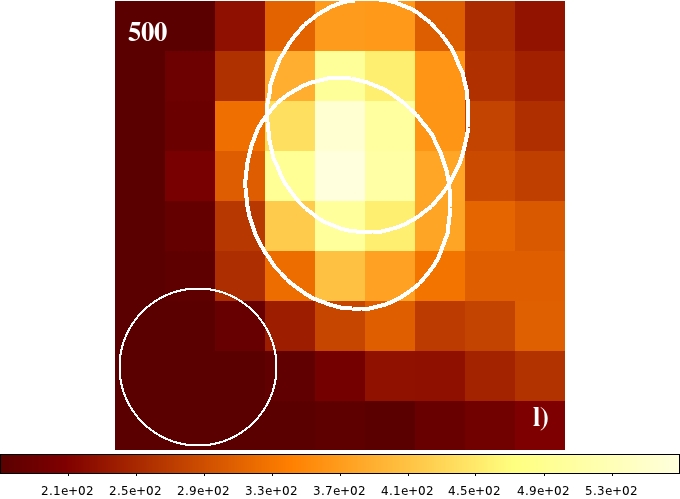}
\includegraphics[width=0.24\linewidth]{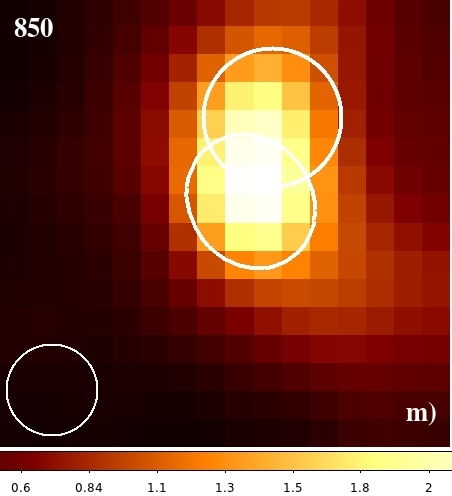}
\includegraphics[width=0.24\linewidth]{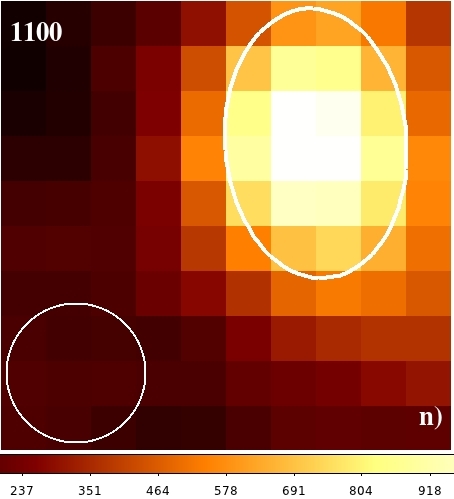}
\caption{Maps of 1\farcm5$\times$1\farcm5 of B1-bS and B1-bN at all wavelengths, with sizes and positions derived from the Gaussian fits. The FWHM of the respective instrumental beam is shown as a circle in the bottom-left corner of each respective panel. At $\lambda\le100$~$\mu$m, the source S295 is also visible. Colour bars are in Jy/beam for the SCUBA maps, and MJy/sr in all the others. In the \textit{Spitzer} images, positions and sizes of the sources are copied from the 70~$\mu$m map. IRAC maps are in the top row, from left to right: a) 3.8~$\mu$m; b) 4.5~$\mu$m; c) 5.8~$\mu$m; d) 8.0~$\mu$m. Second row: e) \textit{Spitzer} 24~$\mu$m; f) PACS 70~$\mu$m, B1-bN is not really detected and its flux corresponds to a 1~$\sigma$ rms point source; g) PACS 100~$\mu$m; h) PACS 160~$\mu$m, S295 is no longer visible. Third row: i) SPIRE 250~$\mu$m; j) SPIRE 350~$\mu$m; k) SCUBA 450~$\mu$m, where the beam shown is 11\arcsec, see Appendix~A.2; l) SPIRE 500~$\mu$m. Bottom row: m) SCUBA 850~$\mu$m; n) Bolocam 1.1~mm, 
where only one source has been fitted.}
\label{mappette2}
\end{figure*}

\textit{Spitzer} source S295 is very close to B1-bS. It is visible in \textit{Herschel} maps at $\lambda\le100$~$\mu$m. S295 is commonly associated with the millimeter-wavelength source Bolo~81 \citep{enoch}. Since B1-bS is close to the wings of the \textit{Spitzer} PSF of S295, it could be that B1-bS is not seen at 24~$\mu$m because of its proximity to the bright S295. To test this possibility, we used DAOPHOT on the \textit{Spitzer} map to perform PSF photometry of S295, a method only possible due to the low background level at 24~$\mu$m. B1-bS remains undetected also after S295 is subtracted; from the analysis of the detections, we concluded that any source with a flux $\ge$0.2~mJy, corresponding to a $\ge$5\,$\sigma$ detection, should have been detected, and so we assumed 0.2~mJy flux as an upper limit at 24~$\mu$m for both B1-bS/bN. At shorter wavelengths, the distance between the B1-b sources and S295 is large enough and the background is so low and smooth, that to estimate an upper limit for the fluxes in the IRAC bands we just computed the rms of the background over a region of 11$\times$11 pixels, centred at the 160~$\mu$m positions. We used this value as the 1\,$\sigma$ upper limit for the peak flux which, along with the instrumental FWHM, can then be used to derive an upper limit at the IRAC wavelengths for a 2D Gaussian profile. The upper limits for B1-bS are systematically larger than those of B1-bN. Since the former is close to the brighter S295, it is possible that the B1-bS upper limits suffer some flux leakage from S295. For this work, however, shifting the upper limits by some percentage does not impact the fitting result (see below). The full set of measured fluxes is reported in Table~\ref{photData}. Image cut-outs at all wavelengths centred on B1-bS, 1\farcm5$\times$1\farcm5 in size, are shown in Fig.~\ref{mappette2}.
%in the PACS/SPIRE bands centred on B1-bS, 1\farcm5$\times$1\farcm5 in size, are shown in Fig.~\ref{mappette}; Fig.~\ref{mappette2} shows the images at all wavelengths.

\begin{figure}
\centering
\includegraphics[width=\linewidth]{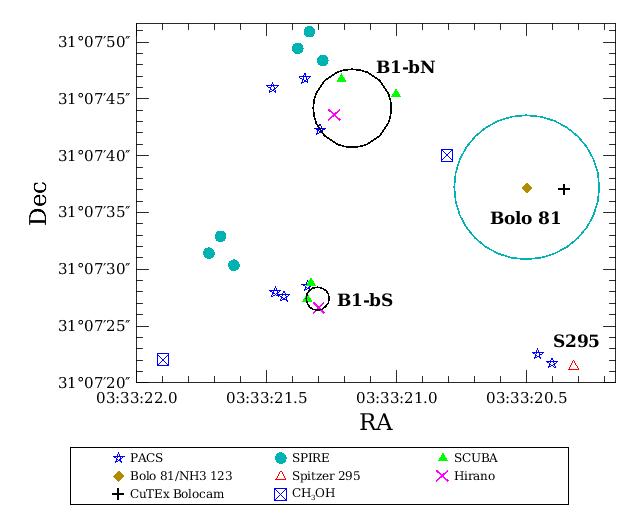}
\caption{Positions (J2000) of the sources at all wavelengths; see text for references. Two black open circles are centred on the position of B1-bS and B1-bN after applying a positional offset correction found by comparing the PACS and \textit{Spitzer} positions of S295, see text for details. The radius of the circles is the 1\,$\sigma$ of the mean of the PACS coordinates, for B1-bS, and of the mean of PACS 100~$\mu$m, 160~$\mu$m, and SPIRE 250~$\mu$m coordinates for B1-bN. The positions at 350~$\mu$m, 500~$\mu$m, and 850~$\mu$m are very uncertain, because B1-bS and B1-bN are not resolved at those wavelengths. The blue open circle centred on Bolo~81 has a radius of 7\arcsec, the pointing accuracy of Bolocam observations \citep{enoch}.}
\label{posizioni}
\end{figure}

The positions of the B1-b sources from 24~$\mu$m to 1.1~mm are shown in Fig.~\ref{posizioni}. The triangle denotes the position of S295 taken from the \textit{Spitzer} c2d catalog \citep{evans}, which agrees with our S295 positions at 70~$\mu$m and 100~$\mu$m. The positions of the B1-b sources given in \citet{hirano} are precessed to J2000. The position of Bolo~81 is from \citet{enoch}. This source was later detected by \citet{roso} in a survey of ammonia cores, and dubbed NH3SRC~123. The cross labelled CuTEx Bolocam is the position we find using CuTEx on the 1.1~mm map. We also report in the figure the positions of the ``core'' and ``outflow'' recently found by \citet{oberg} from CH$_3$OH observations. The position they name ``core'' is halfway between Bolo~81 and B1-bN, while their ``outflow'' is slightly SE of B1-bS.

To derive the \textit{Herschel} coordinates of the B1-b sources we proceeded as follows. First, we averaged the PACS positions of S295 and compared them with the coordinates reported in the c2d catalog, finding an offset of $\Delta\alpha=1\farcs4$ and $\Delta\delta=0\farcs7$. Assuming that the c2d coordinates are more accurate, these offsets can be considered estimates of the absolute value of the PACS pointing errors in our maps. Next, we averaged the PACS coordinates of B1-bS after adding the offset found for S295, finding $\alpha=3^\mathrm{h}$33$^\mathrm{m}$21\farcs3, $\delta=+31^\mathrm{d}$7\arcmin27\farcs4, with a total uncertainty (1\,$\sigma$) of 1\farcs1. We did not average the SPIRE positions because at 350~$\mu$m and 500~$\mu$m the two sources are more blended and the Gaussian fits are less reliable. We used the SPIRE 250~$\mu$m position to estimate the offset between PACS and SPIRE: $\Delta\alpha=4\farcs2$ and $\Delta\delta=3\farcs0$. Finally, for B1-bN, we applied to the PACS 100~$\mu$m 
and 160~$\mu$m coordinates the same offsets found for S295; and to the SPIRE position at 250~$\mu$m the offsets found between PACS and SPIRE for B1-bS. The three pairs of coordinates were averaged finding $\alpha=3^\mathrm{h}$33$^\mathrm{m}$21\farcs2, $\delta=+31^\mathrm{d}$7\arcmin44\farcs2, with a total uncertainty (1\,$\sigma$) of 3\farcs7. Note that all the offsets are much smaller than the separations between the objects, so that sources confusion due to a mis-pointing is excluded.

Even if it was already noted in the past that S295 and B1-bS are spatially separated, in the literature there is some tendency to confuse the two sources. From Figs.~\ref{mappette2} and \ref{posizioni}, it is clear that these two objects are different. The association of Bolo~81 with S295 is very dubious since the latter is not detected longward of 100~$\mu$m. The large offset, $\sim$20\arcsec, between Bolo~81 and the B1-b sources also casts some doubts on the possible association between these objects.

%The distance between Bolo~81 and B1-bS is $\sim16$\arcsec, larger than the pointing accuracy of Bolocam observations of 7\arcsec\ \citep{enoch}. Also the association between these two source, then, is doubtful.

\section{Results of SED fitting}
The simplest model that can be fitted to the data is an optically thin, isothermal greybody, which is appropriate for starless and pre-stellar cores
\begin{displaymath}
F_\nu = \frac{\kappa_\nu B(T,\nu)}{D^2}M
\end{displaymath}
where $\kappa_\nu$ is the opacity (see the discussion in Appendix~A.3), $B(T,\nu)$ is a blackbody at temperature $T$, $D$ is the distance, $M$ is the mass. Following the approach used for other GBS obervations \citep[e.g.,][]{vera}, the dust opacity index was fixed to 2. The best-fit models, considering only data at $\lambda\ge160$~$\mu$m, have $T=9$~K, $M=7.3$~$M_\odot$ for B1-bS, and $T=8$~K, $M=9.4$~$M_\odot$ for B1-bN.

We had to exclude from the greybody fitting the fluxes at short wavelengths because in a two-colour diagram of [100-160] vs. [250-350], the two B1-b sources have colours that are bluer than those of a blackbody. Such a colour could only be reproduced with a greybody given an unrealistic situation where the dust opacity increases with (far-IR/submm) wavelength. If the entire SEDs are fitted with a greybody, as in Equation~(\ref{GB}) and leaving all the parameters free to vary, the best-fit models have $\beta = 0$, and $T=18$~K for B1-bS; and $\beta=0$ and $T=15$~K, for B1-bN. During the fitting procedure, we imposed the condition $\beta\ge0$, and the value $\beta=0$ best matches the condition $\beta<0$, corresponding to a dust opacity increasing with wavelength.

Another way of fitting an SED with colours that are bluer than those of a blackbody is by assuming that the observed emission is due to two components \citep[see, e.g., Fig.~2 in][]{pezzuto}. For this reason, we fitted the B1-b derived SEDs by adding the contributions of a blackbody embedded in a dusty envelope whose emission is modelled with a greybody. We give a physical interpretation of this two-component model later on in this section.

\begin{figure*}
%\sidecaption
\centering
\includegraphics[width=0.45\linewidth]{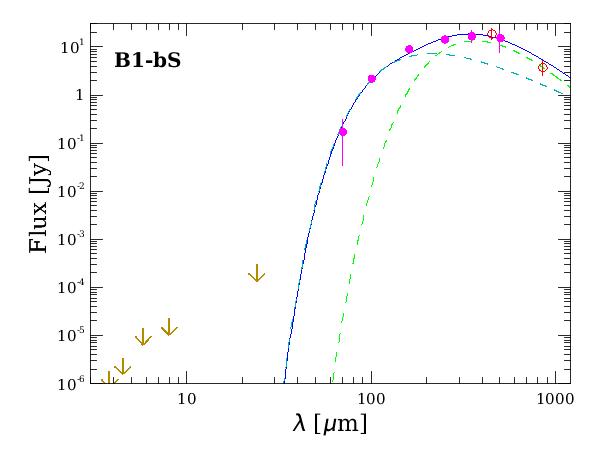}
\includegraphics[width=0.45\linewidth]{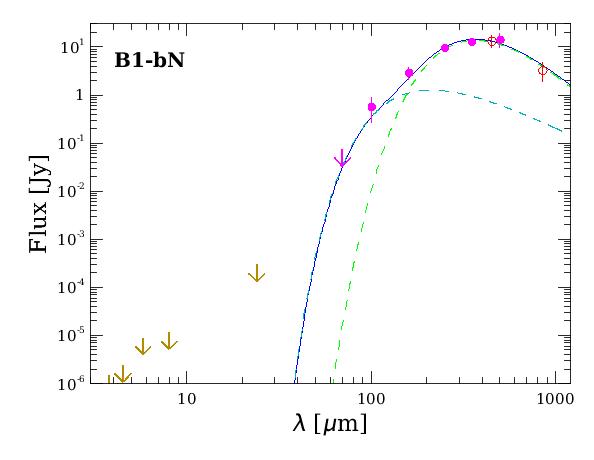}
\caption{The best-fit two-component models for B1-bS and B1-bN. Filled circles are \textit{Herschel} data (for B1-bN the 70~$\mu$m flux is an upper limit), open circles are SCUBA data, and upper limits at $\lambda\le24$~$\mu$m are from \textit{Spitzer} IRAC and MIPS. The solid line is the best-fit model, a sum of a blackbody (dashed line dominating at shorter wavelengths) plus a greybody (dashed line more important at longer wavelengths). \textit{Herschel} fluxes are those reported in Table~\ref{photData} multiplied by the colour correction factors, see Appendix~A.1.}
\label{bfit}
\end{figure*}

The details of the fitting procedure, as well as some additional information on colour corrections applied to the PACS and SPIRE data, are given in Appendix~A. The best result of the two-component fitting procedure is shown in Fig.~\ref{bfit} for both B1-b sources, and the corresponding parameters of the best-fit models, $T_b$ and $\Omega_b$ for the blackbody component, and $T_g$, $\lambda_0$, $\beta$, and $\Omega_g$ for the greybody component, are listed in Table~\ref{results}.

%\begin{table*}\tiny
\begin{table}\tiny
\caption[]{%Left section: t
The parameters of the best-fit models. $T_b$ and $\Omega_b$ are the temperature and the solid angle of the blackbody; $T_g$, $\lambda_0$, $\beta$ and $\Omega_g$ are the parameters of the greybody (see Appendix~A). $\Omega_b$ and $\Omega_g$ are given as radii $(\sqrt{\Omega/\pi})$. Uncertainties are 1\,$\sigma$. The $\chi^2$ are 5.2 and 4.0 for B1-bS and B1-bN, respectively. For B1-bN, the best-fit model has the smallest $\Omega_b$, so we do not have an estimate for the lower uncertainty.\label{results}}% Right section: masses and luminosities derived from the best-fit parameters. $M_\mathrm{env}$ is the mass of the dusty envelope from Equation~(\ref{newMass}); $m_\mathrm{BE}$ is the Bonnor-Ebert mass from Equation~(\ref{mBE}); $L_b$ and $L_g$ are the luminosities of the blackbody and of the greybody components, respectively, the latter found by numerical integration.\label{results}}% The best-fit model found by fitting one component only to each source is a greybody with $\beta = 0$, i.e., a blackbody, with $T=18$~K, $\chi^2=8.0$, for B1-bS; and $T=15$~K, $\chi^2=8.2$, for B1-bN.\label{results}}
%\begin{tabular}{cccccccc|ccccc}%p{0.5\linewidth}l}
\begin{tabular}{ccccccc}%p{0.5\linewidth}l}
\hline
%\noalign{\smallskip}
%Source&$T_b$&$\Omega_b$&$T_g$&$\lambda_0$&$\beta$&$\Omega_g$&$\chi^2$&$M_\mathrm{env}$&$m_\mathrm{BE}$&$L_\mathrm{bol}$&$L_\mathrm{350}/L_\mathrm{bol}$&$T_\mathrm{bol}$\\
%&(K)&($10^{-11}$sr)&(K)&($\mu$m)&&(\arcsec)&&($M_\odot$)&($M_\odot$)&($L_\odot$)&(\%)&K\\
&$T_b$&$\Omega_b$&$T_g$&$\lambda_0$&$\beta$&$\Omega_g$\\
B1-&(K)&(\arcsec/10)&(K)&($\mu$m)&&(\arcsec)\\
%\noalign{\smallskip}
\hline
\noalign{\smallskip}
%B1-bS&29.0$^{+2.5}_{-3.5}$&2.04$^{+0.49}_{-0.39}$&8.0$\pm$0.5&100$^{+15}_{-25}$&2.00$\pm$0.25&20.4$^{+5.3}_{-1.9}$&5.2&9.0$^{+5.6}_{-5.1}$&0.366$^{+0.097}_{-0.040}$&0.48&25&18\\
%B1-bN&33.0$^{+1.5}_{-8.5}$&0.28$^{+0.27}_{-0.00}$&8.0$\pm$0.5&120$\pm$45&2.00$\pm$0.25&17$^{+11}_{-4}$&4.0&9$^{+13}_{-8}$&0.30$^{+0.19}_{-0.07}$&0.28&35&14\\
bS&29.0$^{+2.5}_{-3.5}$&5.25$^{+0.63}_{-0.50}$&8.0$\pm$0.5&100$^{+15}_{-25}$&2.00$\pm$0.25&20.4$^{+5.3}_{-1.9}$\\
bN&33.0$^{+1.5}_{-8.5}$&1.94$^{+0.94}$&8.0$\pm$0.5&120$\pm$45&2.00$\pm$0.25&17$^{+11}_{-4}$\\
\noalign{\smallskip}
\hline
\end{tabular}
\end{table}
%\end{table*}

The best-fit blackbody has a temperature of $T\sim30$~K in both sources. Its size is about 130~AU at 235~pc for B1-bS, and is much smaller in B1-bN. For the latter, $\Omega_b$ is not well constrained since we do not have a 70~$\mu$m detection. The parameters of the greybody component, i.e., the dusty envelope, are similar for both sources. The temperature is 8~K, not too much different from the 11.7~K temperature found by \citet{roso} for the ammonia core NH3SRC~123 detected at the position of Bolo~81. % (see Fig.~\ref{maps}).

From the relation $\dot{m}=\alpha c^3_\mathrm{s}/G$, where $c_\mathrm{s}$ is the sound speed, we can estimate the mass accretion rate. There are two limiting cases for the gravitational collapse of an isothermal sphere \citep[see][for a detailed discussion and references]{mckee}: in the \textit{fast} collapse (the so-called Larson-Penston-Hunter solution) the infall is supersonic and $\alpha=47$, while in the \textit{slow} collapse (or Shu's solution) the infall is sonic and $\alpha=0.975$. For $T=8$~K we derive in the former case $\dot{m}=5.3\times10^{-5}$~$M_\odot/$yr, and $\dot{m}=1.1\times10^{-6}$~$M_\odot/$yr for the latter case. Taking into account that $c_\mathrm{s}$ could include also contributions from a magnetic field or from turbulence, we can safely conclude that $\dot{m}<1\times10^{-4}$~$M_\odot/$yr.

Turning these mass accretion rates into an accretion luminosity is not immediate because we need to know the mass and the radius of the accreting source. The fact that we model the emission of the central object with a blackbody rules out the possibility to derive its mass. As far as the radius is concerned, the value we found from the fitting procedure appears larger than a few AU expected for an FHSC: 130~AU for B1-bS and 48~AU, but with a 50\% error, for B1-bN.

Such a large radius, however, can have a physical meaning. \citet{bate} found that in a rotating envelope a disc with a size of $\sim 100$~AU can form before the stellar core; in the \citet{sato}'s model, again for a rotating core, 100~AU correspond to the radius where the envelope's density profile changes from a $r^{-2}$ power law to a flatten profile which ends on the core's surface, located at 20~AU when the FHSC first forms. The inner component of our model, then, can be physically interpreted as describing the thicker and inner regions of the envelope that eventually merge with the compact central objects. In other words, the parameters $T_b$ and $R_b$ of the blackbody gives the average properties of this \textit{hybrid} component, without being able to distinguish where the envelope ends and where the core starts.

The best-fit value for $\beta$, the exponent of the dust opacity, is 2, even if good fits for B1-bN can be obtained also with $\beta=1.5$; see Appendix~A. The wavelength at which $\tau=1$, $\lambda_0$, is 100~$\mu$m for B1-bS and a value slightly larger for B1-bN.

The mass of the dusty envelope in B1-bS is $9\pm5$~$M_\odot$, with the large uncertainty mainly due to the uncertainty in $\beta$. The mass of B1-bN corresponding to the best-fit model is also 9~$M_\odot$, but the large uncertainty in $\Omega_b$ and $\Omega_g$ causes a large range of variation for the mass: $1\le M/M_\odot\le23$. Previous estimates of the masses were made by \citet{hirano} who, with data at $\lambda\ge350$~$\mu$m, derived for both sources smaller masses, $M\sim1.7$~$M_\odot$, and higher temperatures, $T\sim18$~K. \citet{hatchell} derived an upper limit for B1-bS $M_{\mathrm{235pc}}<23.1$~$M_\odot$, but using for the SED also the millimetre flux of Bolo~81.

The radius $\sqrt{\Omega_g/\pi}$ of the envelope is $\sim$20\arcsec\ for B1-bS and $\sim$17\arcsec\ for B1-bN. Such a radius, however, can not be immediately compared with the sizes given as FWHM in Table~\ref{photData}, which result from fitting the sources in the map with a 2D-Gaussian profile. In fact, when fitting the SED we assume that the source has a finite radius, $R$, while a Gaussian profile does not have any radius. To compare $R$ with the observed FWHMs, we proceeded as follows: the integral $I$ of a normalized 2D circular Gaussian over a circle of radius $R_I$ is \citep{woerz}
\begin{displaymath}
I=1-\exp^{-\frac{R_I^2}{2\sigma^2}}
\end{displaymath}
Clearly, $I=1$ only for $R_I\rightarrow\infty$, but already when $R_I=3\sigma$ $I\sim0.99$\footnote{More precisely, $I=0.9889...$, slightly smaller than 0.9973, the integral over the interval $\pm3\sigma$ for a 1D-Gaussian.}. We then define the radius $R_{\mathrm{G}}$ of a 2D Gaussian as $R_{\mathrm{G}}=3\sigma$, or, using the relation between FWHM and $\sigma$, $R_{\mathrm{G}}=1.29\times$FWHM. For B1-bS, the best-fit radius $R$ is 20\farcs4, a circular 2D-Gaussian with $R_{\mathrm{G}}=R$ has FWHM = 15\farcs8, a value in between the measured size at 250~$\mu$m and at 350~$\mu$m. For B1-bN, the best-fit radius is 17\arcsec, which corresponds to FWHM = 13\arcsec, very close to the measured size at 250~$\mu$m.

It is evident from Table~\ref{photData}, however, that the size derived from the observations is not well defined, since the FWHM increases with wavelength. This trend is to some extent related to the instrumental beam size. For instance, the FWHM of B1-bS is smaller at 100~$\mu$m than at 70~$\mu$m, which is likely related to the smaller PSF of the intrument at 100~$\mu$m (scanning speed of 20\arcsec/s, nominal PACS compression mode) than at 70~$\mu$m (scanning speed of 60\arcsec/s, double PACS compression mode). This trend is also likely to be related to the increase with wavelength of the background level, which makes it more difficult to disentangle the genuine source emission from the extended emission. It is physically plausible, however, to assume that this trend is, at least in part, intrinsic to the sources, given the fact that this phenomenon has been already observed elsewhere.

In high-mass star-forming regions, for instance, an increase of the size of the dense cores with wavelength is known and is taken into account by scaling the flux proportionally to the ratio of measured radii, assuming the size at 160~$\mu$m as the fiducial radius \citep[see, e.g.,][]{hobys,vela}. This approach is justified by theoretical models showing that the emission can be described as coming from an almost isothermal envelope whose mass increases linearly with the radius.

For cores in nearby star-forming regions, where the spatial resolution is high enough to resolve the structure of the envelope, the wavelength-radius relationship could be a consequence of temperature stratification in the envelope. Namely, at shorter wavelengths, we see the inner and warmer part of the envelope. At larger wavelengths, however, the emission comes from the outer and generally colder part of the core. In our model, part of the blackbody radiation is absorbed by the envelope which should cause a temperature stratification. We did not however treat the radiative transfer of the system, assuming instead, for simplicity, an isothermal envelope. The temperature $T_g$ so derived, and in general the whole set of parameters $T_g,\beta,\lambda_0,\Omega_g$, along with the mass $M$ from Equation~(\ref{newMass}), can then be considered to be the averaged properties of the outer regions of the envelope, in the same way as $T_b$ and $R_b$ describe average properties of the inner regions as discussed above. In other words, while our objects likely have a temperature stratification which causes an inrease of the observed radius with wavelength, our model shows that the SEDs can be described in a simple way as having one radius and one temperature at all wavelengths.

The consistency of our approach can be seen from the fact that: 1) the derived size is comparable to the observed sizes at $\lambda\sim300$~$\mu$m, and 2) $\lambda_0\ga100$~$\mu$m. In fact, in a greybody model, the dependence on the solid angle disappears in the optically thin regime, so that the derived $\Omega_g$ values are most likely to represent the sizes corresponding to the largest wavelength where the envelope is not yet completely thin. The robustness of our result can be verified by noting that $T_g$ and $M$ are very similar to the values found with the isothermal greybody model discussed at the beginning of this section.

Having modelled the observed SED with two components allows us to derive separately the internal luminosity $L_{\mathrm{int}}$ of the sources and the bolometric luminosity $L_{\mathrm{bol}}$.

In Table~\ref{lum} we report: $L_b$, the luminosity of the blackbody component. This is found with the standard equation $L_b=4\pi R_b^2\sigma T_b^4$ (the parameters are those of Table~\ref{results} with $R_b=(\sqrt{\Omega_b/\pi})$). At any frequency $\nu$, the envelope absorbs a part of the blackbody emission leaving the amount $B_\nu(T_b)\mathrm{e}^{-\tau_\nu}$ free to escape from the envelope. The integral over frequency of this radiation is reported in the table as $L_\mathrm{em}$. Clearly, the fraction $(1-\mathrm{e}^{-\tau_\nu})$ is absorbed inside the envelope. The column labelled $L_g$ reports the luminosity of the greybody component found by numerical integration of Equation~(\ref{GB}) in the range $1\,\mu\mathrm{m}\le\lambda\le10$~mm. The observable luminosity predicted by the model is then $L_\mathrm{em}+L_g$, which we report in the column $L_{\mathrm{bol}}$. It can be compared with the observed luminosity that we report in the column $L_{\mathrm{SED}}$, found by integrating the measured fluxes.

\begin{table}%\tiny
\caption[]{%Left section: t
The luminosities of our sources, all in $L_\odot$. $L_b$ is the luminosity of the blackbody, which in our model corresponds to the so-called internal luminosity; $L_\mathrm{em}$ is the amount of blackbody radiation that escapes the envelope, i.e., $\int B_\nu(T_b)\mathrm{e}^{-\tau_\nu}\mathrm{d}\nu$; $L_g$ is the luminosity of the greybody component; $L_{\mathrm{bol}}$ is the sum of $L_\mathrm{em}$ and $L_g$, the total observable luminosity predicted by the model; $L_{\mathrm{SED}}$ is the observed luminosity found by integrating the measured fluxes.\label{lum}} \begin{tabular}{cccccc}%p{0.5\linewidth}l}
\hline
%\noalign{\smallskip}
%Source&$T_b$&$\Omega_b$&$T_g$&$\lambda_0$&$\beta$&$\Omega_g$&$\chi^2$&$M_\mathrm{env}$&$m_\mathrm{BE}$&$L_\mathrm{bol}$&$L_\mathrm{350}/L_\mathrm{bol}$&$T_\mathrm{bol}$\\
%&(K)&($10^{-11}$sr)&(K)&($\mu$m)&&(\arcsec)&&($M_\odot$)&($M_\odot$)&($L_\odot$)&(\%)&K\\
B1-&$L_b$&$L_\mathrm{em}$&$L_g$&$L_{\mathrm{bol}}$&$L_{\mathrm{SED}}$\\
%\noalign{\smallskip}
\hline
\noalign{\smallskip}
%B1-bS&29.0$^{+2.5}_{-3.5}$&2.04$^{+0.49}_{-0.39}$&8.0$\pm$0.5&100$^{+15}_{-25}$&2.00$\pm$0.25&20.4$^{+5.3}_{-1.9}$&5.2&9.0$^{+5.6}_{-5.1}$&0.366$^{+0.097}_{-0.040}$&0.48&25&18\\
%B1-bN&33.0$^{+1.5}_{-8.5}$&0.28$^{+0.27}_{-0.00}$&8.0$\pm$0.5&120$\pm$45&2.00$\pm$0.25&17$^{+11}_{-4}$&4.0&9$^{+13}_{-8}$&0.30$^{+0.19}_{-0.07}$&0.28&35&14\\
bS&0.45&0.26&0.22&0.48&0.49\\
bN&0.11&0.04&0.22&0.26&0.28\\
\noalign{\smallskip}
\hline
\end{tabular}
\end{table}

If there is no external source of energy, then by the conservation of energy $L_g=L_b-L_\mathrm{em}$. The contribution of the interstellar radiation field to the sources luminosity could be, then, estimated as $L_{\mathrm{ISRF}}=L_g-(L_b-L_\mathrm{em})$. For the two sources, however, we find quite different results: $L_{\mathrm{ISRF}}=0.03\,L_\odot$ for B1-bS and 0.15~$L_\odot$ for B1-bN. While a small difference between $L_{\mathrm{ISRF}}$ in the two sources could be reasonable, such a large discrepancy is more likely due to the uncertanties in the best-fit parameters, in turn due to the uncertanties in the measured fluxes, or to the consequence of the non-proper treatment of the radiation transfer in our model.

Finally, as an example of a model more rigorously treating the radiative transfer, we fitted the SEDs using the on-line fitting tool by Robitaille et al. (2007), that compares the observed fluxes with a grid of theoretical SEDs. This comparison is important also because in this way we can test the hypothesis that our sources are actually more evolved than an FHSC. The models by \citet{rob} are indeed appropriate to describe the emission from an evolved young stellar object (YSO), where a central star already formed. On the contrary, their grid does not contain models that are appropriate for very young sources like an FHSC. We found that no model can account for both the upper limits in the \textit{Spitzer} bands and the far-IR/submm fluxes. The SEDs of our sources are not compatible with any of the 200\,000 models of YSOs in the grid of \citet{rob}.

%\begin{figure}
%\sidecaption
%\centering
%\includegraphics[width=0.4\linewidth]{/home/stefano/scienza/PACS_perseus/publications/perseus4/lettera/B1-S.jpg}
%\includegraphics[width=0.4\linewidth]{/home/stefano/scienza/PACS_perseus/publications/perseus4/lettera/B1-N.jpg}
%\includegraphics[width=\linewidth]{robFit.jpg}
%\caption{The best-fit model for B1-bS and B1-bN found in the grid by \citet{rob}. The best-fit model happens to be the same for both sources: it has number \#3019992 seen from an angle of 83$^\circ$. Main parameters are $M=0.19$~$M_\odot$, $R=4.11$~$R_\odot$, $T=3040$~K, and an age of $\sim20\,000$~yrs. The envelope outer radius is 3320~AU. The filled circles are the SED of B1-bS, and the crosses, with the upper limit at 70~$\mu$m, are the SED for B1-bN. The solid line shows the fluxes found by integrating over the entire extension of the envelope the emission of the best-fit model. The x's show the fluxes of the best-fit model integrated inside the apertures reported in Table~\ref{photData}. The difference between the solid line and the x's is always negligible, and it is appreciable only between 70~$\mu$m and 250~$\mu$m. The upper limit at 70~$\mu$m has been given a confidence of 68\% because it corresponds to a 1~$\sigma$ level in the noise. For such a confidence level, the minimization procedure allows the model to be brighter than the upper limit. At 24~$\mu$m, the upper limit has been given a confidence of 100\%, being in this case a 5~$\sigma$ limit.}
%\label{robFit}
%\end{figure}

\section{Discussion}
The following features can be extracted from the observed SEDs of our sources, prior to any modelling: the colours in the PACS bands are not compatible with those of a greybody, while the SEDs of the starless or pre-stellar cores can be modelled with a greybody. An additional blackbody is required, which hints at the presence of an inner and warmer compact component. This second component is not visible at $\lambda\la70$~$\mu$m, so it is not as evolved as other nearby protostellar objects. A further observational signature, although indirect, of the very young age of B1-bS comes from the statistics of the YSOs detected in Perseus. The preliminary analysis with CuTEx of the Western region of Perseus gives a list of $\sim40$ tentative sources detected in all of the six \textit{Herschel} bands. All of them are visible in the 24~$\mu$m \textit{Spitzer} map, the only exception being B1-bS. This fact alone points to some peculiarity with this source.

In Table~\ref{compFlux}, we report for all the FHSC candidates the photometric data available in literature, or derived by us, in the fundamental range between 24~$\mu$m and 160~$\mu$m. As already stated in Section~1, the SED only is not enough to identify an FHSC. The other sources reported in Table~\ref{compFlux} have been proposed as FHSC candidates based on other observational properties, e.g., the u-v visibilities from interferometric observations. On the other hand it is evident that as far as the SED is concerned B1-bS remains an exception also when compared with the other candidates. It is the only source that shows a clear evidence of an SED more evolved than a pre-stellar core, but less evolved than a Class~0 source.

{\tiny
\begin{table}
\caption[]{A comparison of the observed SEDs between 24~$\mu$m and 160~$\mu$m for the proposed FHSC candidates. MMSI is in Chamaeleon, CB~17 is a dark globule in the constellation of Camelopardalis, and all the other sources are in the Perseus star-forming region. PACS fluxes and upper limits are from this work with the exception of the PACS 100~$\mu$m flux of CB~17, quoted in \citet{chenF}. \textit{Spitzer} data are from the given reference with the exception of the 24~$\mu$m upper limit for B1-bS and B1-bN, derived by us. For Bolo~45, \citet{schnee} report that this source was not detected in the \textit{Spitzer} map, but no upper limit is given.\label{compFlux}}
\begin{tabular}{l|cccc}%p{0.5\linewidth}l}
\hline
%\noalign{\smallskip}
&24~$\mu$m&70~$\mu$m&100~$\mu$m&160~$\mu$m\\
&\tiny Spitzer&\tiny PACS/Spitzer&\tiny PACS&\tiny PACS/Spitzer\\
Source&(mJy)&(mJy)&(mJy)&(Jy)\\
%\noalign{\smallskip}
\hline
%\noalign{\smallskip}
MMSI\tablefootmark{a}             &2.5   & \phantom{---}/200   &     &\\
CB 17\tablefootmark{b}            &$<$11 &  &36   & \phantom{---}/0.8   \\
Per 58\tablefootmark{c}           &0.88  &67/65    &407  &1.58/2.87  \\
L1451-m\tablefootmark{d}          &$<$1.5&$<$44/$<$72 &$<$15&0.54/0.88  \\
Bolo 45\tablefootmark{e}          &NA    &$<$71/NA    &$<$20&3.96/NA    \\
\tiny L1448-IRS2E\tablefootmark{f}&$<$18 &$<$45/$<$120&165  &3.32/$<$2.7\\
B1-bS\tablefootmark{g}            &$<$0.2&220/\phantom{---}&2290 &9.1/\phantom{---}      \\
B1-bN\tablefootmark{g}            &$<$0.2&$<$50/\phantom{---}      &610  &3.24/\phantom{---}      \\
%\noalign{\smallskip}
\hline
\end{tabular}
\tablebib{\tiny \tablefoottext{a} \citet{belloche}; \tablefoottext{b} \citet{chenF}; \tablefoottext{c} \citet{fhsc}; \tablefoottext{d} \citet{pineda}; \tablefoottext{e} \citet{schnee}; \tablefoottext{f} \citet{chenF2}; \tablefoottext{g} this work.}
\end{table}
}

Other age indicators directly derived from the SEDs are the ratio $L_\mathrm{350}/L_\mathrm{bol}$, i.e., the luminosity at $\lambda\ge350$~$\mu$m over the bolometric luminosity, and $T_\mathrm{bol}$, i.e., the temperature of a blackbody having the same mean frequency as the observed SED. For a Class~0 source, $L_\mathrm{350}/L_\mathrm{bol}>5$\% \citep{class0} and $T_\mathrm{bol}<70$~K \citep{chen}. Our sources have values quite extreme with respect to these limits, thus confirming that they are indeed very young. B1-bS has $L_\mathrm{350}/L_\mathrm{bol}=25$\% and $T_\mathrm{bol}=18$~K; B1-bN has $L_\mathrm{350}/L_\mathrm{bol}=35$\% and $T_\mathrm{bol}=14$~K. From these signatures alone, however, it is not possible to distinguish an FHSC from a pre-stellar core, or a starless core.

In summary, B1-bS shows all the features expected in the SED of an FHSC. B1-bN, the companion 20\arcsec\ north of B1-bS, is not detected at 70~$\mu$m, but otherwise it shares all of the characteristics of B1-bS.

For both sources, only the bolometric luminosities, found by integrating the observed fluxes, do not fit with FHSC predictions: $L_\mathrm{bol}$ is 0.49~$L_\odot$ and 0.28~$L_\odot$ for B1-bS and B1-bN, respectively, in excess of the maximum foreseen for an FHSC, $L\la0.1$~$L_\odot$ \citep{omukai}. Such a high luminosity could be ascribed to imperfect background removal, but even if we limit the integration to the PACS bands, where the diffuse emission is less prominent, we find for B1-bS $L_\mathrm{PACS}=0.14~$~$L_\odot$. It then seems reasonable to expect $L\ga0.2$~$L_\odot$, a value that also excludes the possibility that B1-bS is a very low luminosity objects \citep[VeLLOs; e.g.,][]{vello}.

The observed bolometric luminosity, however, can be larger than the internal luminosity due to external heating of the envelope by the interstellar radiation field. We have seen in the previous section that is not easy to estimate $L_\mathrm{ISRF}$ from our data. Instead, we can make use of the relationship between $L_\mathrm{int}$ and $F_\mathrm{70}$ found by \citet{dunham2}. For B1-bS, their relationship gives $L_\mathrm{int}\sim0.04$~$L_\odot$, a value which seems difficult to reconcile with the observed luminosities. On the other hand, \citet{dunham2} derived the relation by modelling the protostar with a fixed temperature of 3\,000~K and a luminosity in the range of 0.03---10~$L_\odot$, which implies a radius in the range 0.5---11~$R_\odot$. Such values are reasonable for a Class~0 source, but could be inappropriate for a still-younger object, like a FHSC. \citet{dunham2}'s result, however, clearly implies that for our sources $L_\mathrm{int}$ is surely smaller than the measured $L_\mathrm{bol}$. The precise factor, however, remains unknown.

Recently, \citet{comm} found that the luminosity of an FHSC becomes larger than 0.1~$L_\odot$ only at late times, so that the vast majority of FHSCs appear as VeLLOs for most of all their lifetime. Nonetheless, they also found cases where only a few hundred years after the formation of the FHSC, the predicted luminosity is as high as 0.3~$L_\odot$, similar to what we derive for B1-bN. Such an increase in luminosity is due to the increase of mass of the inner core due to the accretion from the envelope.

Another possible contribution to the luminosities of our sources is through contamination of the SEDs by an outflow, as recently found by \citet{maury} for the prototypical Class~0 protostar VLA1623. In our cases, however, the existence of an outflow in B1-bS/bN is not completely clear, so it is not possible to explore the reliability of this hypothesis. \citet{drabek} found that $^{12}$CO line emission can contribute to the dust continuum at 850~$\mu$m. They estimated, however, that in NGC~1333, at positions far from outflows, the contamination is less than 20\%. For the B1 region, Sadavoy et al. (in preparation) found that CO (3-2) emission appears to be relatively minor towards B1-bS and B1-bN, therefore any contamination of the dust continuum from the gas is likely insignificant at 850~$\mu$m.

The results of the fits we performed must be interpreted cautiously given the uncertanties in the fluxes. We tentatively suggest, however, that B1-bN is still slightly younger than B1-bS. It has an envelope slightly more optically thick than that of B1-bS, explaining the lack of detection in the shortest PACS band. For both sources, the envelope is cold and massive, and its internal thermal pressure alone insufficiently high to be able to prevent its gravitational collapse. This conclusion comes from comparing the mass of the envelope, and the corresponding critical Bonnor-Ebert mass $m_\mathrm{BE}$, the largest mass that an isothermal sphere of gas bounded by pressure can have without collapsing. When $M_\mathrm{env}>m_\mathrm{BE}$, the internal thermal pressure may not be high enough to support the core against internal gravity and external pressure, so that the envelope is undergoing, or is about to collapse. We found for $m_\mathrm{BE}$, see Equation~(\ref{mBE}), 0.4~$M_\odot$ (B1-bS) and 0.3~$M_\odot$ (
B1-bN). In both cases, $M_\mathrm{env}\gg m_\mathrm{BE}$ so that the envelopes cannot be supported by internal thermal pressure alone, even if we cannot exclude that other physical mechanisms like the amount of turbulence or the strength of the magnetic field \citep[see, e.g.,][]{basu}, may still contribute to the support of the cloud against the gravitational collapse

As far as the magnetic field is concerned, however, the largest mass that a magnetic field of 31~$\mu$G \citep[as estimated for the B1 region by][]{mawi} can support is only 0.11~$M_\odot$ \citep[from][]{stutz}, assuming a radius of 0.023~pc (20\arcsec at 235~pc, see $\Omega_g$ in Table~\ref{results}).

To estimate the stability against turbulence we computed the virial parameter $\alpha$ \citep{bertoldi} and we consider a core gravitationally bound if
\begin{displaymath}
\alpha=\frac{5\sigma^2R}{GM}\la2
\end{displaymath}
$\sigma$ is the velocity dispersion found by adding in quadrature the non-thermal component of the NH$_3$ lines measured by \citet{roso} for NH3SRC~123, and the thermal component of a mean particle of molecular weight $\mu=2.33$: $\sigma_{\mathrm{NT}}=0.325$~km/s and $\sigma_{\mathrm{T}}=0.204$~km/s. From Table~\ref{results} $R=7.19\times10^{16}$~cm and $R=5.99\times10^{16}$~cm for B1-bS and B1-bN, respectively, for a distance of 235~pc. Then, $\alpha=0.44$ and 0.37 for the two sources. Clearly, the exact values of $\alpha$ are not well constrained given the large uncertainties, especially in the mass of the sources. As long as, however, $M\ga2$~$M_\odot$, $\alpha\la2$.

Important information on the evolutionary stage of a source can be obtained from the observations of its outflow. The conclusions drawn from spectral observations \citep[e.g.,][]{hatchell2,oberg}, however, have been affected by the lack of an adequate knowledge of the spatial distribution of the sources. Recall that \citet{hirano} detected only B1-bS/bN, while in the Bolocam survey at 1.1~mm \citep{enoch} only Bolo~81 was detected. Also, recall that the \textit{Spitzer} maps \citep{rebull} show only S295, while in the SCUBA map at 850~$\mu$m B1-bS and B1-bN are blended. Thanks to \textit{Herschel} data, for the first time we can see clearly B1-bS/bN and S295 in the same map, in particular in the PACS bands. This new definition will allow a better understanding of the spectroscopic observations. For instance, by comparing our Fig.~\ref{posizioni} with the CH$_3$OH maps by \citet{oberg}, it appears possibile that B1-bS coincides with the peak in the CH$_3$OH map, in between the two identified outflows (the 
authors give the coordinates of only one outflow, which we reported in Fig.~\ref{posizioni}).

The presence of outflows makes it possible that one of the two sources is actually a local density enhancement (knot). Indeed, \citet{gueth} concluded that emission knots generated by the shocked outflow in the vicinity of a protostellar object can be misinterpreted as, e.g., starless clumps. This conclusion, however, has been derived from mm and submm maps; to what extent it holds also in the FIR is unknown.

We finally note that the number of FHSC candidates found in Perseus is quite large. \citet{pineda} derived that the number of expected FHSCs in this region should be $\le$0.2, 30 times smaller than the 6 proposed candidates reported in Table~\ref{compFlux}. Beside the obvious argument that eventually none of the 6 candidates may be confirmed to be an FHSC, there are other ways to explain this discrepancy. The expected number of FHSCs is found by assuming that the ratio of the number of FHSCs over the number of Class~0 sources is equal to the ratio between their respective lifetimes. Since the two ratios are equal only in the case of continuous star formation, a first hypothesis is that we are observing a burst of star formation in Perseus. The other possibility, which like the previous one has already been suggested by \citet{pineda}, is that we assume incorrect lifetimes for the FSHC phase, the Class~0 phase, or both. Also the number of Class~0 sources may be underestimated \citep{schnee}, which would also 
impact the estimated lifetime. Though the catalog of detected sources in Perseus is not yet released (Pezzuto et al., in preparation), it is reasonable to suppose that \textit{Herschel} observations will update the number of starless cores, pre-stellar cores and Class~0 sources for all nearby star-forming regions. With this information, it is possible that the expected number of FHSCs will change.

Finally, we note that our sources are quite close, with a projected separation of 20\arcsec. If the physical distance is not very different from the projected one, then B1-bS and B1-bN are $\ga$4\,700~AU apart, corresponding to few Jeans lengths at 10~K. It is then possible that these sources formed more or less at the same time from the fragmentation of a larger structure. This possibility, if confirmed, would explain why we found two FHSC candidates at close proximity.

\section{Conclusions}
In this paper, we have presented the results of fitting the SEDs of two sources in the Perseus star-forming region. Data were derived from \textit{Herschel} photometry observations collected within the Gould Belt Survey program \citep{pandreGB}. The two sources are B1-bS and B1-bN, discovered by \citet{hirano} from interferometric observations at millimeter wavelengths. B1-bS is the only source in the Western part of the star forming region in Perseus, detected in all six \textit{Herschel} bands, and not visible in the \textit{Spitzer} 24~$\mu$m map. These two criteria are important for the SED-based detection of FHSC candidates. The SED alone is not enough to determine the evolutionary status of a source, and other proposed FHSC candidates have been observed with interferometry to better assess their status. But limiting to the photometric criteria B1-bS is the only candidate that satisfies the detection at 70~$\mu$m and the non detection at 24~$\mu$m.

\textit{Herschel} data were complemented with \textit{Spitzer}, JCMT-SCUBA, and CSO-Bolocam data. The resulting SED was fitted with a two-component model that describes the emission in terms of a blackbody, which roughly corresponds to the compact central object, and a greybody. i.e., the surrounding dusty envelope. We also tried to model the SED with a simple greybody alone, which is adequate for a pre-stellar core, and with a proper radiative transfert model \citep{rob}, suited for more evolved sources. Both latter models failed to reproduce the SED over the whole observed spectral range. We conclude that B1-bS shows almost all the expected characteristics expected in an FHSC. Only its luminosity is too high with respect to the theoretical predictions. Indeed, such a high luminosity is at present the strongest argument against our candidate being an FHSC.

The SED of B1-bN is similar to that of B1-bS, with the important difference that the former is not detected at 70~$\mu$m. For the rest, this source seems similar to B1-bS. The two sources are few 10$^3$~AU apart, corresponding to few Jeans lengths. It is then possible that these two sources formed at almost the same time from the fragmentation of a larger structure.

Further observations at long wavelengths with high spatial resolution, such those that ALMA will undertake, are clearly needed to understand better the nature of B1-bS and B1-bN. It is reasonable, however, to conclude that \textit{Herschel} has observed two young objects which could be in a very early stage of protostellar formation, and we propose them as two new FHSC candidates.

\begin{acknowledgements}
We thank the anonymous referee for the many comments and suggestions that made the paper by far better than the original. We thank also B. Commer\c{c}on for helping us to answer one of the points raised by the referee; P. Martin and J. Kirk for helpful clarifications on the dust opacity; A. Men'shchikov for running \textit{getsources} on our map to independently verify our extraction. SP spent one week at the Herzberg Institute of Astrophysics in Victoria, during which this work was finalized; he thanks all the people at the institute for the kind and warm hospitality. KLJR, DP, MP, DE and ES are funded under ASI contracts I/038/08/0 and I/005/11/0.

PACS has been developed by a consortium of institutes led by MPE (Germany) and including UVIE (Austria); KU Leuven, CSL, IMEC (Belgium); CEA, LAM (France); MPIA (Germany); INAF-IFSI/OAA/OAP/OAT, LENS, SISSA (Italy); IAC (Spain). This development has been supported by the funding agencies BMVIT (Austria), ESA-PRODEX (Belgium), CEA/CNES (France), DLR (Germany), ASI/INAF (Italy), and CICYT/MCYT (Spain).
SPIRE has been developed by a consortium of institutes led by Cardiff Univ. (UK) and including: Univ. Lethbridge (Canada); NAOC (China); CEA, LAM (France); IFSI, Univ. Padua (Italy); IAC (Spain); Stockholm Observatory (Sweden); Imperial College London, RAL, UCL-MSSL, UKATC, Univ. Sussex (UK); and Caltech, JPL, NHSC, Univ. Colorado (USA). This development has been supported by national funding agencies: CSA (Canada); NAOC (China); CEA, CNES, CNRS (France); ASI (Italy); MCINN (Spain); SNSB (Sweden); STFC, UKSA (UK); and NASA (USA). HIPE is a joint development by the Herschel SGSC, consisting of ESA, the NASA HSC, and the HIFI, PACS and SPIRE consortia.
\end{acknowledgements}

%\Online
\appendix

\section{SED fitting and photometry}
For our two-component models, we modelled each set of data as the sum of a blackbody and a greybody envelope: $F_\nu=B(T_b)\mathrm{e}^{-\tau}\Omega_b+G(T_g,\lambda_0,\beta)\Omega_g$, where
\begin{equation}
G(T_g,\lambda_0,\beta) = (1-\mathrm{e}^{-\tau})B_\nu(T_g)\Omega_g\label{GB}
\end{equation}
with $\tau$, the optical depth, parametrized as a power-law, i.e.,
\begin{equation}
\tau = \left(\frac{\lambda_0}{\lambda}\right)^\beta\label{tau}
\end{equation}
where $\lambda_0$ is the wavelength at which $\tau = 1$, and $\Omega_g$ and $\Omega_b$ are the solid angles of each respective component. The independent, unknown parameters are then $T_b,T_g,\lambda_0$, and $\beta$ while the solid angles are computed by scaling the model fluxes to the data. This model describes our sources in terms of two components: a central source, the blackbody, embedded in a dusty envelope whose emission is modelled with a greybody. Since we do not impose that the envelope is optically thin at all wavelengths, the blackbody radiation is attenuated by a factor that describes the absorption due to the dust opacity. The absorption term, strictly speaking, implies that the envelope cannot be isothermal, so, as explained in Section~3, the set of parameters $T_g,\lambda_0,\beta,\Omega_g$ should be considered only as describing the average physical conditions of the envelope. The density profile in the cores usually follows a power law, $\rho\propto r^{-q}$, with typically $q\la2$, so that 
the mass, and then the flux, increases with the radius. $T_g$ should be then indicative of the temperature in the outer region of the envelope.

The best-fit model was found inside a grid prepared by varying the parameters in the following intervals: $5\le T(\mathrm{K})\le 50$ in steps of 1~K; $0\le\beta\le 5$ in steps of 0.5; and $10\le\lambda_0(\mu\mathrm{m})\le600$ in steps of 10~$\mu$m\footnote{We took into account the quantization of the parameters in the grid: for instance, all the good models for B1-bS have $\beta=2.0$, and, since the step in $\beta$ is 0.5, we quote the result as $\beta=2.00\pm0.25$.}. These large intervals, larger than physically plausible, were chosen to ensure that the best-fit parameters would not fall on the border of the grid. The best fit was then chosen as the one with the smallest $\chi^2$. During the search, we have applied a few constraints: a) $T_b>T_g$, i.e., the greybody envelope must be colder than the blackbody component; b) $\Omega_b<\Omega_g$, the blackbody must be smaller than the greybody; c) the model must predict a 24~$\mu$m flux smaller than the upper limit; d) the predicted flux at 1.1~mm must be 
smaller than the measured value; and e) the envelope mass must be in the range of 0.1--10~$M_\odot$.

Once the minimum $\chi^2$ was found, the 1\,$\sigma$ uncertainties in the parameters were found by considering all the models with a $\chi^2$ in the range $\chi^2_\mathrm{min}\le\chi^2\le\chi^2_\mathrm{min}+1$ \citep{rene}. For B1-bS, only 8 models were found in this $\chi^2$ range. For B1-bN, not having a 70~$\mu$m flux had the consequence that 242 models, out of a total of 27646, had $\chi^2<\chi^2_\mathrm{min}+1$. Among these 242 models, the best fit was the one reported in Table~\ref{results}, but with very large uncertainties. To decrease the uncertainties, we added another constraint and selected only the 107 models with $\lambda_0\le200$~$\mu$m, in analogy with the results of the selection for B1-bS. Finally, assuming that the dust properties are the same in the two sources, we considered only the models with $\beta=2$, discarding 70 models with $\beta=1.5$.

For the Bonnor-Ebert mass, we used the formula reported in \citet{vera}
\begin{equation}
m_{\mathrm{BE}}=2.4\frac{Ra^2}{G} \label{mBE}
\end{equation}
where $a$ is the sound speed corresponding to the temperature found from the fit, assuming a molecular weight $\mu=2.33\mu_{\mathrm{H}}=3.90\times10^{-24}\,$g. For $R$ we used the radius found from the fit, i.e., $R=D\sqrt{\Omega_g/\pi}$.

\subsection{Photometry corrections}
Before comparing the model fluxes with the observed data, two steps were performed. For the first step, a colour correction was made; the flux calibration of PACS and SPIRE was performed under the usual assumption that the SED of a source displays a flat $\nu F_\nu$ spectrum. For all other kinds of SED, the derived fluxes must be colour corrected according to the intrinsic source spectrum. To correct the fluxes properly, however, we need to know the spectrum a priori but that is actually what we want to derive from the data themselves. To overcome this circular problem, we computed a correction the other way round: a set of greybody models were computed over a large frequency range and the fluxes were derived as
\begin{equation}
F_\mathrm{cc}=\int F_\nu(T,\lambda_0,\beta)\mathrm{RSRF}_\nu\mathrm{d}\nu
\end{equation}
where RSRF$\nu$ stands for the relative spectral response function of each PACS or SPIRE filters, computed by the instruments control teams. To obtain the corresponding flux density, $F_\mathrm{cc}$ was divided by the appropriate filter width that we computed by imposing the condition that the colour corrections at the effective wavelengths is 1 for a SED of constant $\nu F_\nu$. The derived $\Delta\lambda$ and $\Delta\nu$ are reported in Table~\ref{photoSystem}. As a consistency check, we compared the colour corrections we obtained for a blackbody with the values found by the instrument teams (see http://herschel.esac.esa.int/). For PACS, the agreement is better than 1\% in all the bands for $T\ge7$~K. The agreement at lower temperatures is the same at 100~$\mu$m and 160~$\mu$m, while it starts to diverge at 70~$\mu$m, in particular our correction for $T=6$~K is 1.3\% higher, at $T=5$~K is 37\% higher. At such extremly low temperatures, however, the colour corrections are quite difficult to be estimated 
since the intrinsic spectrum differs notably from the calibration spectrum. For SPIRE the comparison is less obvious because the colour corrections for a blackbody are not tabulated as a function of the temperature, but only in the limiting case of a blackbody in the Rayleigh-Jeans regime. For a blackbody at 100~K, we found that our colour correction is less than 1.3\% higher at 250~$\mu$m, and less than 1\% for the other two bands.

\begin{table}%\tiny
\caption[]{Effective wavelengths and filter widths used to compute the colour corrections.\label{photoSystem}}
\begin{tabular}{lcccccc}%p{0.5\linewidth}l}
\hline
%\noalign{\smallskip}
$\lambda$ ($\mu$m)&70&100&160&250&350&500\\\hline
$\Delta\lambda$ ($\mu$m)&10.6&17.0&30.2&46.3&60.5&113\\
$\Delta\nu$ ($10^{11}$Hz)&6.48&5.09&3.54&2.22&1.93&1.35\\
\hline
\end{tabular}
\end{table}

The colour corrections for the attenuated blackbody were found by inverting the definition of the greybody, i.e., $B_\nu(T)\mathrm{e}^{-\tau} = B_\nu(T)-G_\nu(T)$, from which $(B_\nu(T)\mathrm{e}^{-\tau})_\mathrm{cc}=B_\mathrm{cc}(T)-G_\mathrm{cc}(T)$. In this way, the same grid of models could be used.

Once the best fit was found, we computed the true colour correction factors, i.e., the amount by which the observed fluxes must be multiplied to get the instrumental corrected fluxes. For the best fit models reported in Table~\ref{results}, the colour corrections $f_\mathrm{cc}$ are given in Table~\ref{fcc}.

\begin{table}%\tiny
\caption[]{Colour correction factors for the best fit models reported in Table~\ref{results}. \textit{Herschel} fluxes reported in Fig.~\ref{bfit} are the measured fluxes given in Table~\ref{photData} divided by these factors.\label{fcc}}
\begin{tabular}{ccccccc}%p{0.5\linewidth}l}
\hline
%\noalign{\smallskip}
Source&70&100&160&250&350&500\\
\hline
B1-bS & 0.790 & 0.978 & 0.999 & 1.016 & 1.006 & 0.987\\
B1-bN & 0.744 & 0.946 & 0.909 & 1.029 & 1.015 & 0.992\\
\hline
\end{tabular}
\end{table}

For the second step we made before comparing model fluxes with observed fluxes, we took into account the Gaussian fit used to derive the photometric flux. For example, it is known that the PACS PSF is not a Gaussian, so an error is introduced in the photometry because of the Gaussian fit. To derive these errors, we reduced a set of PACS observations of flux calibrators and the results of aperture photometry and of synthetic photometry (Gaussian fitting) were compared. This exercise was repeated for a set of isolated compact sources in the Perseus field with different integrated fluxes. For 70~$\mu$m, 100~$\mu$m, and 160~$\mu$m, we found correction factors of 1.6, 1.5, and 1.4, respectively.\ %For instance, the rms of 14.1 MJy/sr we found at 70~$\mu$m, corresponds to a point source flux of 0.030~Jy (for a beam of $\sim$9\arcsec), which becomes 0.048~Jy after the multiplication by 1.6. 
We are making additional tests to improve our knowledge of Gaussian fit errors. For SPIRE, the PSF is much closer to a Gaussian profile, so that we do not need correction factors for SPIRE fluxes.

In summary, Table~\ref{photData} reports the measured fluxes multiplied by the factors that correct for the Gaussian fit. In Fig.~\ref{bfit}, the \textit{Herschel} data are the fluxes from Table~\ref{photData} multiplied by the color correction. For instance, the flux at 70~$\mu$m of B1-bS resulting from the Gaussian fit is 0.138~Jy, which becomes 0.22~Jy after the multiplication by 1.6. This is the value reported in Table~\ref{photData}. After being multiplied by 0.79, see Table~\ref{fcc}, the flux becomes 0.17~Jy, which is the value used in Fig.~\ref{bfit}.

We do not yet have estimates of the uncertainties in the maps, so it is not possible to give reliable uncertainties to each flux. The uncertainties given in Table~\ref{photData} were found after a comparison of CuTEx fluxes with those found with the \textit{getsources} algorithm \citep{sasha}.

\subsection{Photometry in SCUBA maps}
The effective beam of SCUBA at 450~$\mu$m is a Gaussian of FWHM 17\farcs3 \citep{james}, but the actual beam consists of two different spatial components: an inner one with a FWHM of 11\arcsec, and an outer one with a FWHM of 40\arcsec. The sizes we derived for B1-bS and B1-bN are smaller than 17\farcs3. Since CuTEx is more sensitive to compact sources than to extended sources, we conclude that what CuTEx fitted was just the inner component of the sources, while the extended component, larger than the sources' separation, was seen by CuTEx as background. To estimate the total flux of the individual sources, we used the following approach. If a source has an intrinsic size $\theta$, then
\begin{displaymath}
F_\mathrm{tot}=P_1\left(\theta_1/11\right)^2+P_2\left(\theta_4/40\right)^2
\end{displaymath}
where $\theta_1=\sqrt{\theta^2+11^2}$ and similarly for $\theta_4$. From the analysis of \citet{james}, we know that the peak fluxes are $P_1=0.88P$ and $P_2=0.12P$, where $P$ is the peak flux of the 17\farcs3 FWHM Gaussian, from which $P_2=P_1\times0.12/0.88$. Finally, expressing $\theta_4$ as a function of $\theta_1$, we have
\begin{equation}
F_\mathrm{tot}=P_1\left(\theta_1/11\right)^2+\frac{0.12}{0.88}P_1\left(\frac{\theta_1^2-11^2+40^2}{40^2}\right)\label{f450}
\end{equation}

At 850~$\mu$m, CuTEx finds sizes that are larger than the effective beam of 22\farcs9, but since this beam is again the combination of two Gaussians \citep{james} of 19\farcs5 and 40\arcsec\ FWHM, it is likely that also in this case CuTEx fits just the inner Gaussian and sees the second one as a background. The formula we used to find the total flux is the same as Equation~(\ref{f450}), with 19\farcs5 instead of 11\arcsec. Since our sources are slightly larger than the effective beam, however, the applicability of Equation~(\ref{f450}) to the 850~$\mu$m flux is less robust.

\subsection{The mass of a greybody}
Without \textit{Herschel} data to define the SED in the far-infrared, the emitting mass is usually derived from the flux measured at long wavelengths where the envelope becomes optically thin. In this limit Equation~(\ref{GB}) then becomes
\begin{equation}
F_\nu \approx \tau B_\nu(T)\Omega\label{GBot}
\end{equation}
If the envelope is optically thin, we see the whole mass distribution $M$, so that
\begin{equation}
\tau=\kappa_\nu\int\rho\mathrm{d}s\approx\kappa_{\mathrm{ref}}\left(\frac{\lambda_{\mathrm{ref}}}{\lambda}\right)^\beta\frac{M}{A}\label{taugiusta}
\end{equation}
where $A$ is the projected area and $\kappa_{\mathrm{ref}}$ is the opacity at the reference wavelength $\lambda_{\mathrm{ref}}$. As in other papers of the Gould Belt consortium, e.g., \citet{vera}, we adopted an opacity of $\kappa_\mathrm{ref}=0.1\,\mathrm{cm}^2\mathrm{g}^{-1}$ at $\lambda_\mathrm{ref}=300$~$\mu$m \citep{beck}. With this choice, the GBS opacity law is very similar to the one by \citet{hild}, who adopted $\kappa_\mathrm{ref}=0.1\,\mathrm{cm}^2\mathrm{g}^{-1}$ at $\lambda_\mathrm{ref}=250$~$\mu$m and $\beta=2$ for $\lambda\ge250$~$\mu$m. For this work, actually, only $\kappa_{\mathrm{ref}}$ is important because $\beta$ was derived directly from the fit. $\lambda_{\mathrm{ref}}$ was shifted to 300~$\mu$m for consistency with older results obtained with ground-based facilities, like IRAM or JCMT. By comparing the column density map of IC~5146 obtained from \textit{Herschel} observations and the GBS dust opacity law with that obtained from SCUBA (sub)mm data and with a different 
opacity law, \citet{arzo} found that the GBS dust opacity law  works well for the \textit{Herschel} data. The exact value for $\kappa_\mathrm{ref}$ is in any case uncertain and likely dependent on the dust temperature. An uncertainty in the derived mass of a factor $\sim$2 cannot be excluded.

Substituting Equation~(\ref{taugiusta}) into Equation~(\ref{GBot}) and writing the solid angle as $A/D^2=\pi R^2/D^2$, with $R$ equal to the outer radius of the envelope, assumed spherical, and $D$ equal to the distance to the source, we arrive at the well known formula
\begin{equation}
M \approx  \frac{F_\nu D^2}{\kappa_\nu B_\nu(T)}\label{oldMass}
\end{equation}
Since from our best-fit models $\lambda_0$ and $\beta$ are also found, we derived a different formula to estimate the mass of a greybody envelope.

At wavelengths where the envelope is optically thin, we find by combining Equations~(\ref{tau}) and (\ref{taugiusta}) that
\begin{displaymath}
\left(\frac{\lambda_0}{\lambda}\right)^\beta\approx\kappa_{\mathrm{ref}}\left(\frac{\lambda_{\mathrm{ref}}}{\lambda}\right)^\beta\frac{M}{A}
\end{displaymath}
from which
\begin{equation}
M=\frac{D^2\Omega}{\kappa_{\mathrm{ref}}}\left(\frac{\lambda_0}{\lambda_{\mathrm{ref}}}\right)^\beta\label{newMass}
\end{equation}
This new equation, equivalent for a greybody to the unnumbered equation on page~274 in \citet{hild}, gives the mass inside a sphere of radius $R$ that, for the given opacity law, has optical depth 1 at $\lambda_0$. Of course, Equation~(\ref{newMass}) can be applied only if we know $\beta$ and $\lambda_0$, but these are what \textit{Herschel} data allow us to find. The mass of the envelopes, discussed in Section~3, were found using the above formula where the greybody parameters are those derived from the fit and reported in Table~\ref{results}.

\section{Sigma clipping with a threshold dependent on the sample size}
Before generating the final maps, the bolometers' time series have to be corrected for the presence of glitches caused by the impact of high energy particles falling onto the detectors. To achieve this correction, we exploited the spatial redundancy provided by the telescope movement, which ensures that each sky pixel of the final map has been observed several times with different bolometers. Outlier detection was then made with a sigma-clipping algorithm. Namely, given a sample of $N$ values, estimates for the mean and standard deviations are derived. All the values that differ from the mean by more than $\alpha$ standard deviations are considered outliers and removed from the sample.

One problem with the sigma-clipping algorithm is the choice of the number of standard deviations above which a datum is considered an outlier. To avoid making an arbitrary choice of $\alpha$, a formula has been derived that makes $\alpha$ dependent on the size of the sample. Below, we give some details of the method.

For a Gaussian distribution with mean $m$ and standard deviation $\sigma$, the probability to get a value $x$ such that $m-\alpha\sigma<x<m+\alpha\sigma$, is given by erf($\alpha/\sqrt{2})$ where erf($y$) is the error function
\begin{displaymath}
\mbox{erf}(y)=\frac2{\sqrt{\pi}}\int_0^y\mbox{e}^{-t^2}\mbox{d}t
\end{displaymath}
For example, if $\alpha=3$, the probability is erf($3/\sqrt{2})=0.9973$. Obviously, $1-$erf$(\alpha/\sqrt{2})$ gives the probability to get a value outside the same interval% $m-\alpha\sigma\le x\le m+\alpha\sigma$
. This probability tends to zero when $\alpha\rightarrow\infty$ (by definition erf($+\infty$)=1).

In a sample of $N$ data, the probability given by the error function, multiplied by $N$, can be interpreted as a the expected number of points inside or outside a certain interval around the mean. For any value of $\alpha$, the number $p$ of points that differ more than $\alpha\sigma$ from the mean is then
\begin{displaymath}
p(\alpha;N)=\left[1-\mbox{erf}\left(\frac{\alpha}{\sqrt{2}}\right)\right]\cdot N
\end{displaymath}
For an ideal Gaussian, any value of $\alpha$ is allowed. In a real experiment, however, $p$ is an integer number and data differing from the mean by $\alpha\sigma$, such that $p(\alpha;N)\ll1$, are suspicious. To implement this condition in the sigma-clipping algorithm, we have to define a precise value of $\alpha$, say $\tilde{\alpha}$, above which observed data are considered outliers and removed from the sample. To this aim, we define $\tilde{\alpha}$ as
\begin{displaymath}
p(\tilde{\alpha};N)=1
\end{displaymath}
or
\begin{equation}
\mbox{erf}\left(\frac{\tilde{\alpha}}{\sqrt{2}}\right)=1-\frac1N\label{erfN}
\end{equation}

All the points, if any, that are outside the interval $m-\tilde{\alpha}\sigma\le x\le m+\tilde{\alpha}\sigma$ are considered outliers and removed from the sample. This is equivalent to assume that $p(\alpha>\tilde{\alpha};N)\equiv0$ instead of $p(\alpha>\tilde{\alpha};N)<1$.

When $N\rightarrow\infty$ then $\tilde{\alpha}\rightarrow\infty$ too. The mathematical property of the Gaussian distribution that for an infinitely large sample any value is allowed is preserved.

\begin{figure}
\centering
\includegraphics[width=0.8\linewidth]{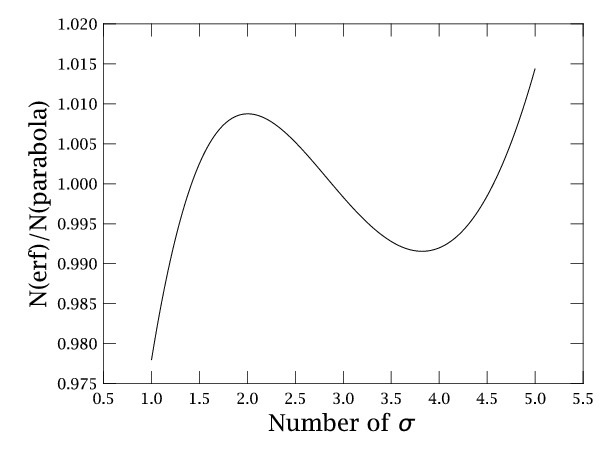}
\caption{In this figure we show, as a function of the number of standard deviations, the ratio between $N\equiv N(\tilde{\alpha})$ as given by Equation~(\ref{erfN}), and $N\equiv N(\tilde{\alpha})$ approximated with the parabola given in Equation~(\ref{parabola}). The difference between the two curves is always less than 3\%. Fitting a cubic instead of a parabola does not improve significantly the approximation and makes the relation $N\equiv N(\tilde{\alpha})$ more nuisance to invert.\label{n_sigma}}
\end{figure}

It is not possible to invert analytically Equation~(\ref{erfN}) for a given $N$, but it is possible to derive an approximate analytical expression. Indeed, we found that in the range $1\le\tilde{\alpha}\le5$ Equation~(\ref{erfN}), written in terms of $\log(N)$, can be approximated with the parabola
\begin{equation}
\log(N) = 0.0794+0.2282\tilde{\alpha}+0.2005\tilde{\alpha}^2\label{parabola}
\end{equation}
In Fig.~\ref{n_sigma} we show the ratio between $N$ as given by Equation~(\ref{erfN}) and by Equation~(\ref{parabola}). The ratio is always less than 3\%.

Inverting Equation~(\ref{parabola}) is trivial, the result is
\begin{equation}
\tilde{\alpha}=-0.569+\sqrt{-0.072+4.99\log(N)}\label{nfromN}
\end{equation}

For a given $N$, the value of $\tilde{\alpha}$ is found from the above equation. The number of expected points distant from the mean by more than $\tilde{\alpha}\sigma$ is zero. If one or more outliers are instead found, they are removed. New values of $m$ and $\sigma$ are recomputed and the new $N^\prime<N$ is used in Equation~(\ref{nfromN}). This process is repeated until no other outliers are present in the sample. The procedure may not converge, for instance, if the noise is not Gaussian. For this reason, we do not actually iterate the formula. Instead, for each point of the map the detection of outliers is done only once. The number of outliers found at the first iteration, however, can be larger than 1.

In the central region of the map, where the coverage (and thus $N$) is high, the number of standard deviations for clipping is larger than in the outskirts of the map where the coverage is low. For instance, if a sky pixel has been observed with 40 bolometers, the above formula gives $\tilde{\alpha}=2.25$. So, once we have estimated the mean $m$ and the standard deviations $\sigma$, all of the values $x_i$ such that ABS$(x_i-m)>2.25\sigma$ are flagged as outliers. If instead a pixel has been observed with 20 bolometers the threshold lowers to 1.96$\sigma$.
\end{document}